%% file: 0.main.tex
\documentclass[acmsmall]{acmart}
\AtBeginDocument{%
  \providecommand\BibTeX{{%
    \normalfont B\kern-0.5em{\scshape i\kern-0.25em b}\kern-0.8em\TeX}}}

\setcopyright{acmcopyright}
\copyrightyear{2023}
\acmYear{2023}

\acmJournal{JACM}
\acmVolume{37}
\acmNumber{4}
\acmArticle{111}
\acmMonth{8}

\usepackage{enumitem}
\usepackage{tabu}
\usepackage{multirow}
\usepackage{graphicx}
\usepackage{hhline}
\usepackage{amsmath, bm}
\usepackage{subfigure}
\usepackage{algorithm}
\usepackage{algpseudocode}
\usepackage{hyperref}
\usepackage{boldline}

\newcommand{\para}[1]{{\vspace{2pt} \bf \noindent #1 \hspace{0.5pt}}}

\newenvironment{packed_itemize}{
\vspace{-\topsep}
\begin{list}{\labelitemi}{\leftmargin=1.5em}
  \setlength{\itemsep}{1pt}
  \setlength{\parskip}{0pt}
  \setlength{\parsep}{0pt}
  \setlength{\headsep}{0pt}
  \setlength{\topskip}{0pt}
  \setlength{\topmargin}{0pt}
  \setlength{\topsep}{0pt}
  \setlength{\partopsep}{0pt}
}{\end{list}\vspace{-\topsep}}

\theoremstyle{definition}
\newtheorem{definition}{Definition}[section]

\setcopyright{none}
\acmJournal{TOIS}
\begin{document}

\title{Privacy-Preserving Individual-Level COVID-19 Infection Prediction via Federated Graph Learning}

\author{Wenjie Fu}
\email{wjfu99@outlook.com}
\affiliation{%
  \institution{Research Center of 6G Mobile Communications, School of Cyber Science and Engineering, and Wuhan National
Laboratory for Optoelectronics, Huazhong University of Science and Technology}
  \city{Wuhan}
  \country{China}
}

\author{Huandong Wang}
\email{wanghuandong@tsinghua.edu.cn}
\author{Chen Gao}
\email{chgao96@tsinghua.edu.cn}
\affiliation{%
  \institution{Beijing National Research Center for Information Science and Technology (BNRist),
Department of Electronic Engineering, Tsinghua University}
  \city{Beijing}
  \country{China}}

\author{Guanghua Liu}
\email{guanghualiu@hust.edu.cn}
\affiliation{%
  \institution{Research Center of 6G Mobile Communications, School of Cyber Science and Engineering, Huazhong University of Science and Technology}
  \city{Wuhan}
  \country{China}
}

\author{Yong Li}
\email{liyong07@tsinghua.edu.cn}
\affiliation{%
  \institution{Beijing National Research Center for Information Science and Technology (BNRist),
Department of Electronic Engineering, Tsinghua University}
  \city{Beijing}
  \country{China}}

\author{Tao Jiang}
\email{taojiang@hust.edu.cn}
\affiliation{%
  \institution{Research Center of 6G Mobile Communications, School of Cyber Science and Engineering, Huazhong University of Science and Technology}
  \city{Wuhan}
  \country{China}
}

\renewcommand{\shortauthors}{Fu, et al.}

\begin{abstract}
Accurately predicting individual-level infection state is of great value since its essential role in reducing the damage of the epidemic. However, there exists an inescapable risk of privacy leakage in the fine-grained user mobility trajectories required by individual-level infection prediction. In this paper, we focus on developing a framework of privacy-preserving individual-level infection prediction based on federated learning (FL) and graph neural networks (GNN). We propose \emph{Falcon}, a \textbf{F}ederated gr\textbf{A}ph \textbf{L}earning method for privacy-preserving individual-level infe\textbf{C}tion predicti\textbf{ON}.
It utilizes a novel hypergraph structure with spatio-temporal hyperedges to describe the complex interactions between individuals and locations in the contagion process.
By organically combining the FL framework with hypergraph neural networks, the information propagation process of the graph machine learning is able to be divided into two stages distributed on the server and the clients, respectively, so as to effectively protect user privacy while transmitting high-level information. Furthermore, it elaborately designs a differential privacy perturbation mechanism as well as a plausible pseudo location generation approach to preserve user privacy in the graph structure. Besides, it introduces a cooperative coupling mechanism between the individual-level prediction model and an additional region-level model to mitigate the detrimental impacts caused by the injected obfuscation mechanisms. Extensive experimental results show that our methodology outperforms state-of-the-art algorithms and is able to protect user privacy against actual privacy attacks. Our code and datasets are available at the link: \url{https://github.com/wjfu99/FL-epidemic}.
\end{abstract}



\maketitle

\input{1.introduction}

\input{3.preliminary}

\input{4.method}
\input{5.evaluation}
\input{2.relatedwork}

\section{Conclusion}\label{sec::conclusion}
In this paper, we investigate the individual-level infection prediction for more precise individual-level intervention strategies (e.g., early warning and mobility control) and propose \emph{Falcon}, a privacy-preserving federated graph learning framework. In this framework, we introduce a novel hypergraph construction method to capture the spatio-temporal dependency, then utilize the spatio-temporal hyperedge as a bridge for cross-client information sharing under privacy constraints. Besides, we design two obfuscation mechanisms that strictly guarantee that the users' actual location will not be divulged to the honest-but-curious server. Furthermore, a novel cooperative coupling mechanism is designed to integrate the microscopic model with a macroscopic model for overcoming the performance decline caused by obfuscation mechanisms. Extensive and multi-scenario experiments are conducted for comprehensive results, which verifies that our method outperforms existing state-of-the-art baseline methods. Our profound analysis provides valuable instruction for future research on individual-level infection prediction.


\bibliographystyle{ACM-Reference-Format}
\bibliography{0.main}

\end{document}

%% file: 1.introduction.tex
\section{Introduction}\label{par:intro}

The rapid spread of COVID-19 all over the world has caused great damage to human health and social economy, which creates a strong motivation for us to have an accurate individual-level infection prediction. Specifically, strategies including early warning~\cite{sheehan2020early} and mobility control~\cite{feng2022precise} can be accurately implemented to effectively reduce the damage of the epidemic. However, accurate individual-level infection prediction requires fine-grained user mobility trajectories to characterize the contagion process driven by close contacts between individuals, where there exists growing privacy concerns~\cite{ienca2020responsible, veliz2021privacy}. Since sensitive information such as where each individual went and who he met is contained in his trajectory, which may be leaked in the process. Therefore, how to simultaneously provide accurate individual-level infection prediction and user privacy protection is a research problem of great value.

Since the contagion process of the epidemic is essentially driven by close contacts between individuals~\cite{salathe2010high}, this process is naturally suitable to be characterized and modeled in the form of graph learning~\cite{xia2021graph}. In light of the great success of the graph neural network (GNN) in various fields, e.g., recommendation~\cite{guo2023poincare, yang2021legalgnn, huang2023position, chen2020tgcn}, drug design~\cite{jimenez2020drug,hsieh2021drug} and spatio-temporal prediction~\cite{zhuang2022uncertainty,rao2022graph}, there have been numerous methods based on GNN that have been proposed to implement COVID-19 forecasting~\cite{sesti2021integrating,murphy2021deep}. However, due to the risks of privacy leakage in individual mobility trajectories, most existing methods alternatively implement infection prediction at the regional aggregation level~\cite{sesti2021integrating,cao2022mepognn, zhuang2022uncertainty, kapoor2020examining}, giving up fine-grained individual infection prediction, which also limits their effectiveness in guiding fine-grained epidemic control strategies. Other approaches simply ignore the risk of user privacy leakage~\cite{murphy2021deep,chen2022deeptrace}. At the same time, federated learning (FL) is a distributed machine learning framework, the concept of which is proposed by Google~\cite{konevcny2016federated}. Its goal is to train machine learning models based on training data distributed across multiple devices without the requirement of uploading the data, which only shares the aggregated locally calculated intermediate results between servers and devices. FL techniques provide us with a promising solution to the problem of privacy-preserving individual-level infection prediction.

However, realizing privacy-preserving individual-level infection prediction based on GNN and FL, i.e., federated graph machine learning (FGML), is also an intractable task with the following three distinct challenges.
First, in the context of FGML, there is a risk of client-side subgraph structure leakage when various information follows the propagation mechanism of GNN and is transmitted between the client and the server. 
Specifically, the graph structure utilized in the infection prediction task either characterizes the contact between individuals~\cite{murphy2021deep} or the interaction between individuals and locations~\cite{hall2022supporting}, which may reveal the sensitive information of users to the honest-but-curious server or other users.
For instance, the interaction between individuals and locations can be indirectly exposed to the server by inference attack via non-zero gradient~\cite{melis2019exploiting} or sequential location queries~\cite{shokri2011quantifying}.
Thus, how to characterize the complicated interactions between individuals and locations while protecting user privacy in the graph-structure data is the first challenge. 
Moreover, if we directly implement individual-level infection prediction with vanilla GNN-based models in the manner of FL, it will lead to the second challenge: cross-client missing information~\cite{fu2022federated}. Specifically, each individual conceals its interaction data with other individuals or locations in its local device to avoid the leakage of graph structure, which means each client only owns a subgraph of the global contact graph. 
Besides, clients will try to avoid exposing raw node features on their local subgraph to others~\cite{zhang2021subgraph,fu2022federated}.
Therefore, a client can only aggregate features on its subgraph but cannot access the node features located on other clients' subgraphs, which leads to insufficient node representations~\cite{liu2022federated}.
Finally, various obfuscated mechanisms utilized in FGML guarantee user privacy from multi-perspectives, such as the perturbation mechanism used in FL that can achieve differential privacy~\cite{wei2020federated,huang2021evaluating,yang2022accuracy}. As a double-edged sword, they also result in a decline in model prediction performance due to obfuscation on data quality or structure. Then, how to overcome this performance decline is the third challenge.

In this paper, we propose \emph{Falcon}, a \textbf{F}ederated gr\textbf{A}ph \textbf{L}earning method for privacy-preserving individual-level infe\textbf{C}tion predicti\textbf{ON}.
\emph{Falcon} introduces a novel hypergraph structure with spatio-temporal hyperedge to characterize the complicated interaction between individuals and locations in the contagion process. 
The spatio-temporal hyperedge is employed as the mediator to propagate individual information, where its features are gathered on the server based on intermediate results shared in FL and then transmitted back to the client devices.
By utilizing spatio-temporal hyperedge as the bridge for transferring features of high-order neighbors, Falcon is able to successfully overcome the challenge of cross-client missing information.
Further, \emph{Falcon} is embedded with a novel plausible pseudo location generation technique and a differential private perturbation mechanism, which respectively confront the leakage of graph structure caused by non-zero gradient and sequential location queries in the transfer intermediate results.
Finally, we elaborately design a novel cooperative coupling mechanism, which incorporates the individual-level prediction model with an auxiliary region-level infection prediction model to counteract the performance decline caused by the injected obfuscation mechanisms.  
Overall, our main contributions are summarized as follows:
\begin{packed_itemize}
    \item We design a hypergraph construction method to extract the spatio-temporal interaction among regions and individuals. Then, we detach the propagation procedures of hypergraph into client and server sides, which realizes the cross-client information exchange in a secure manner.
    \item We propose a novel plausible pseudo location generation technique and a differential privacy perturbation mechanism against two kinds of location inference attacks from the honest-but-curious server. They use the lowest possible cost of prediction utility to exchange for user privacy.
    \item We design a novel cooperative coupling mechanism between the macroscopic region-level model and the microscopic individual-level model, which is able to overcome the performance decline caused by the utilized obfuscation mechanisms.
    \item Extensive and multi-scenario experimental results show that our proposed framework is able to implement accurate infection prediction, outperforming state-of-the-art algorithms. In addition, it can protect user privacy against actual privacy attacks.
\end{packed_itemize}

The structure of this paper is as follows. We first define the problem and introduce preliminary knowledge in Section~\ref{sec::pre}. We then present our solution in Section~\ref{sec::method} and conduct experiments in Section~\ref{sec::exp}. Last, we review the related works in Section~\ref{sec::related} and conclude the paper in Section~\ref{sec::conclusion}.

%% file: 3.preliminary.tex
\section{Preliminaries}\label{sec::pre}
In this section, we provide an overview of our paper by introducing fundamental concepts and formally delineating the infection prediction task at the individual level that we investigate. In addition, a brief review of hypergraph neural networks (HGNN) is conducted as follows. The key notations utilized in this paper are described in Table~\ref{tab:notation}.

\begin{table}[htbp]
\footnotesize
    \begin{center}
        \caption{Key notations used in this paper.}\label{tab:notation}
        \begin{tabular}{m{2cm}<{\centering}|m{10cm}}
            \hlineB{3}
            \textbf{Notation} & \textbf{Description} \\ \hline
            $\mathcal{N}$ & The number of individuals.\\ \hline
            $\mathcal{N}_{r,t}$ & The number of individuals located in the region $r$ at time $t$.\\ \hline
            $\mathcal{M}$ & The number of regions.\\ \hline
            $\mathcal{M}_{u,t}$ & The number of regions visited by the user $u$ at time $t$.\\ \hline
             $L_u$ & The historical trajectory of user $u$.\\ \hline
             $y_u$ & The infection status of user $u$.\\ \hline
             $\mathbf{\Theta}$ & The parameters of the prediction model.\\ \hline
             $\mathcal{G}$ & The spatio-temporal hypergraph .\\ \hline
             $\mathcal{V}$ & The vertex set of hypergraph.\\   \hline
             $\mathcal{E}$ & The hyperedge set of hypergraph.\\\hline
             $\mathcal{H}$ & The adjacent matrix of hypergraph.\\ \hline
             $\mathbf{D}_v$ & The degree matrix of vertex.\\ \hline
             $\mathbf{D}_e$ & The degree matrix of hyperedge.\\ \hline
             $\mathbf{E}^{l+1}$ & The aggregate hyperedge embedding.\\ \hline
             $\mathbf{E}_m$ & The output hidden state of the macroscopic model.\\ \hline
             $\sigma_l$ & The standard deviation of the Gaussian noise for the perturbation mechanism.\\ \hline
             $\sigma_f$ & The standard deviation of the Gaussian noise for DPSGD.\\ \hline
             $\mathcal{G}^{(t)}$ & The graph of the macroscopic model at time $t$.\\ \hline
             $\boldsymbol{X}^{(t)}$ & The new cases of each region observed at time $t$.\\ \hline
             $\beta$ & The transmission rate of the disease.\\ \hline
             $\mu$ & The recovery rate of the infected individual.\\ \hline
             $1/\alpha$ & The average duration of the latent period.\\ \hline
             $R_0$ & The basic reproductive number of a disease. \\ 
            \hlineB{3}
        \end{tabular}
        \label{tab:symbal}
    \end{center}
\end{table}

\subsection{Problem Formulation}

In the real world, predicting the infection status of each individual is beneficial for the precise design of individual-level intervention strategies. Meanwhile, the spread of the epidemic is driven by individual contacts, which are further characterized by the mobility pattern of each individual. Thus, we conclude this problem as a semi-supervised classification problem, utilizing the mobility data and observed positive cases to predict the health status of the rest of the individuals.
Nevertheless, the mobility data of massive individuals raises several practical issues, including significant computational requirements and broad privacy implications. Transferring the whole task to the manner of FL is an effective way to mitigate these anxieties~\cite{li2020federated}. 

We consider the COVID-19 disease transmission within a city or a district, which involves $\mathcal{N}$ individuals and $\mathcal{M}$ regions. 
The infection status of individuals can be divided into four categories~\cite{yang2020modified}: susceptible, asymptomatic infectious, symptomatic infectious, and recovered. 
An individual under the asymptomatic infectious or symptomatic infectious is considered as an infected case. Based on the government's policies and individual intuitions, we legitimately assume that a proportion $\lambda$ of the total population will go to the hospital or the Center for Disease Control and Prevention (CDC) to obtain their infection status. Note that these institutions provide individuals with accurate infection status diagnosis results through medical techniques, such as molecular tests, antigen tests, etc. Besides, with the rapid development of mobile localization techniques, we reasonably suppose that each individual is capable to record his historical trajectory and always retains the latest $l$ days trajectory. 

As depicted in Figure 1, the overall prediction model is based on the FL framework, in which each individual is considered as a client, and the server is located in a specific agency (e.g., a government agency, the CDC, or a technology corporation). The infection diagnosis result and the historical trajectory are both stored on the local device.

We intend to provide policymakers with a GCN-based individual-level infection prediction model, based on which they can develop more precise individual-level intervention strategies. In the meantime, this model should ensure that each individual's historical trajectory $L_n$ and diagnosis result $y_n$ are still stored on their local device and cannot be accessed by other individuals or the server.

Based on the above notations, we formulate the formal definition of the individual-level COVID-19 infection prediction in the manner of FL as follows:
\begin{definition}[Individual-level COVID-19 Infection Prediction in the manner of FL]
Give a city with $\mathcal{M}$ areas, $\mathcal{N}$ individuals, the reported historical trajectory of all individuals, and the diagnostic results for individuals in the population with a proportion $\lambda$ of the population. The overall goal of the entire FL framework is to train an effective distributed individual-level COVID-19 infection prediction model to predict the infection status of all remaining individuals $\mathcal{N}_{1-\lambda}$:
\begin{equation}
    \left\{\hat{y_u}\right\}_{u \in \mathcal{N}_{1-\lambda}} = \hat{f}(\{L_u^\eta\}_{u \in \mathcal{N}}, \left\{{y_u}\right\}_{u \in \mathcal{N}_{\lambda}}; \mathbf{\Theta}),
\end{equation}
\end{definition}
where $\eta$ represents the proportion of the uploaded trajectory to the complete trajectory. Thus, the lower the value of $\eta$, the lower the frequency of trajectory uploads by the individuals.


\subsection{Hypergraph Neural Networks}\label{Hypergraph Convolution}
Hypergraph is a variant of graph, where an edge can connect with any number of vertices. A hypergraph can be formally expressed as $\mathcal{G} = \left(\mathcal{V}, \mathcal{E} \right)$ in which $\mathcal{V}$ and $\mathcal{E}$ denote sets of vertex and hyperedge, respectively. The adjacent matrix of a hypergraph $\mathcal{G}$ can be represented as follows:
\begin{equation}
    \mathbf{H}\left(i_v, i_e\right) = 
    \left\{
         \begin{array}{lr}
         1, & v\in e, \\
         0, & v\notin e, \\
         \end{array}
    \right.
\end{equation}
where $i_v$ and $i_e$ are the indexes of the vertex and hyperedge, respectively.
Vertex and edge degrees can be defined as $d\left(v\right) = \sum_{e\in\mathcal{E}} h(v,e)$ and $d\left(e\right) = \sum_{v\in\mathcal{V}} h(v,e)$, respectively. Further, $\mathbf{D_e}$ and $\mathbf{D_v}$ denote the diagonal matrices of the edge degrees and the vertex degrees. 

HGNN ~\cite{feng2019hypergraph} extended the graph convolution from graph into hypergraph to capture the high-order correlation, which can be represented as:

\begin{equation}\label{equ:HGNN}
\mathbf{X}^{(l+1)}= \sigma 
\left( 
\mathbf{D}_v^{-1} \mathbf{H} \mathbf{W} \mathbf{D}_e^{-1} \mathbf{H}^\top \mathbf{X}^{(l)} \mathbf{\Theta}^{(l)}
\right),
\end{equation}
where $\mathbf{X}^{(l)} \in \mathbb{R}^{|\mathcal{V}| \times F_i}$ represents the input to the HGNN layer, $\mathbf{\Theta}^{(l)}\in \mathbb{R}^{F_i \times F_o}$ denotes the trainable parameters in the $l$-th layer, $\mathbf{H}^\top$ is the operator to aggregate features from nodes to edges, $\mathbf{H}$ is the operator to transform edge features to nodes. $\mathbf{D}_v$ and $\mathbf{D}_e$, the degree matrices of edge and vertex, are utilized as normalization factors. $\mathbf{W}$ is the weight matrix of the hyperedge.


%% file: 4.method.tex
\section{Methodology}\label{sec::method}


\subsection{System Overview}\label{par:sys}
\begin{figure}[t!]
    \centering
    \includegraphics[width=.95\textwidth]{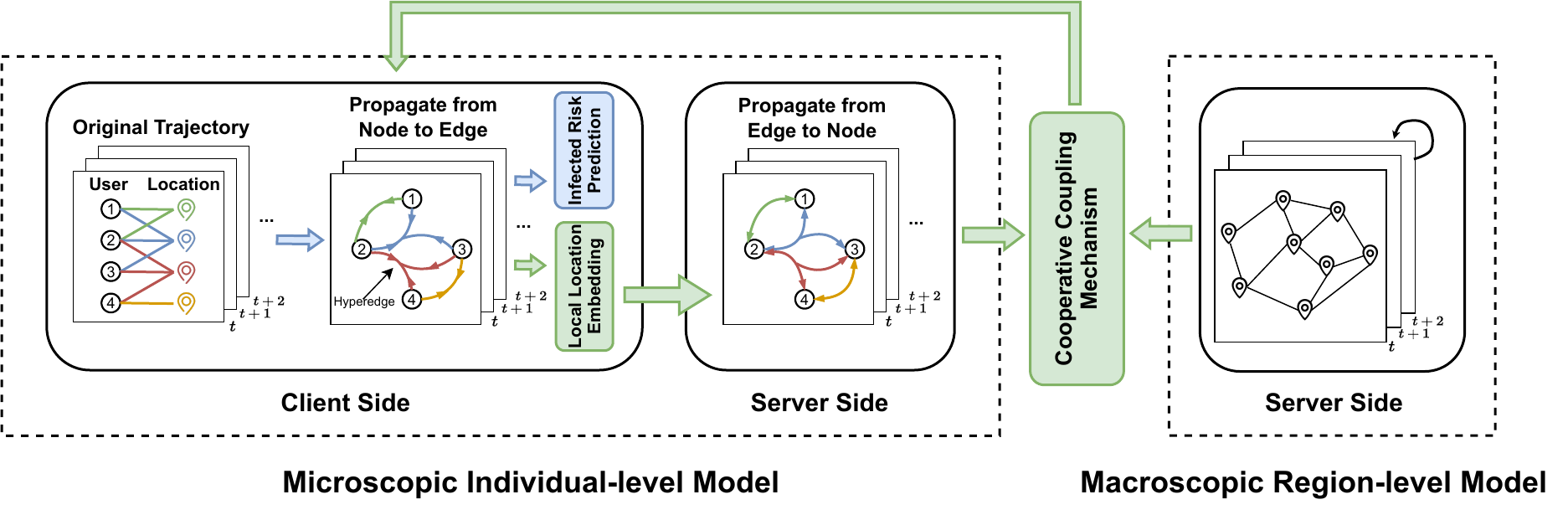}
    \caption{Overall architecture of \emph{Falcon}.}
    \label{fig:overview}
\end{figure}
To tackle the challenges of FGML as mentioned in Section~\ref{par:intro}, we propose a novel algorithm named \emph{Falcon}. The overall architecture of \emph{Falcon} is presented in Figure \ref{fig:overview}, where we use the hypergraph to construct spatio-temporal user-location interaction, then divide two aggregation phases into client and server sides. Besides, a plausible pseudo location generation method is designed against location inference attacks. Furthermore, to eradicate the performance decline caused by the privacy mechanisms, we elaborately design a GNN-LSTM-based macroscopic model to incorporate with our microscopic individual-level model. We summarize the detailed algorithm of \textit{Flacon} in Algorithm \ref{alg:server} and Algorithm \ref{alg:client}.

\subsection{Privacy-preserving FGML Framework}\label{par:privacy mechanism}
First, to extract complex interaction among massive users, we design a spatio-temporal hypergraph construction method as presented in the left of Figure \ref{fig:overview}. Specifically, By regarding each individual as a node, we connect all nodes who visit a location $r$ in the same time interval $t$ to a hyperedge $e(t,r)$, which enables the model to capture the interaction between individuals and locations and the contact among individuals. Consequently, people moving between two spatio-temporal points are represented as the node connects to two distinct hyperedges.

Then, we craft a novel detached hypergraph propagation mechanism in Section~\ref{par:detach} to address the cross-client missing information problem in FGML, which detaches the two phases of hypergraph aggregation to be executed on the client and server, respectively. 
Additionally, we also propose two strictly coupled obfuscation mechanisms, the plausible pseudo location generation algorithm and the differential privacy perturbation mechanism in Section~\ref{par:pseudo} and Section~\ref{par:dp-per-mec}, respectively, to avoid divulging users' actual locations to the server.
In this way, \emph{Falcon} can effectively reduce the privacy leakage risk of the graph structure. Finally, we utilize a differential privacy mechanism~\cite{abadi2016deep} to perturb the gradients of model parameters before the client uploads them to the central server, which can avoid exposing feature values via gradients.

    \begin{algorithm}
    \algrenewcommand\algorithmicrequire{\textbf{Input:}}
\algrenewcommand\algorithmicensure{\textbf{Output:}}
    \caption{Training algorithm of \textit{Flacon} on the server side.}\label{alg:server} 
    \begin{flushleft}
        \textbf{Input:} Initial model parameters of HGNN $\mathbf{\Theta}^0$, initial client and location embeddings  $e_{u,t}^0, e_{r,t}^0$, number of training epochs $N$, learning rate $\eta_{lr}$.\\
        \textbf{Ouput:} Trained global weights ${\theta}_g^{N}$, trained client embedding $e_u^n$.
\end{flushleft}
    \begin{algorithmic}[1]

    \Statex \hspace{-17pt}\textbf{Server executes:}
    \For{epoch $n$ = 1, 2, $\cdots$, $N$}
        \State Distribute $\mathbf{\Theta}^{n-1}$ and $e_{r,t}^{n-1}$ to user clients
        \For{each user client $ u \in \mathcal{N}$} 
        \State $\{e_{u, r,t}^n\}_{r=1}^{n_p \times \mathcal{M}_{u,t}} \leftarrow \textbf{EmbeddingCal}(\{e_{r,t}^{n-1}\}_{r=1}^{\mathcal{M}_{u,t}})$ \hfill \Comment{Update the local location embedding.}
        \EndFor
        \For{each location $r \in \mathcal{M}$}
        \State $    e_{r,t}^n \leftarrow \left( \sum_{u \in \mathcal{N}_{r,t}} e_{u,r,t}^n - (\mathcal{N}_{r,t} - 1) e_{r,t}^{n-1} \right)$
                     \hfill \Comment{Aggregate from nodes to hyperedge.}
        \EndFor
        
        \For{each user client $ u \in \mathcal{N}$}
            \State $\mathbf{g}_u \leftarrow $ \textbf{ClientUpdate}($\{e_{r,t}^n\}_{r=1}^{\mathcal{M}_{u,t}}$) 
        \EndFor
        \State $\mathbf{\Theta}^{n} \leftarrow \mathbf{\Theta}^{n-1} - \eta_{lr} \sum_{u =1}^{\mathcal{N}} \frac{1}{\mathcal{N}} \mathbf{g}_u$ \Comment{Update the parameters of model.}
    \EndFor
            
    \end{algorithmic}
    \end{algorithm}

    \begin{algorithm}
    \algrenewcommand\algorithmicrequire{\textbf{Input:}}
\algrenewcommand\algorithmicensure{\textbf{Output:}}
    \caption{Training algorithm of \textit{Flacon} on the client side.}\label{alg:client} 
    \begin{algorithmic}[1]
        \Statex \hspace{-17pt} $\textbf{EmbeddingCal}(\{e_{r,t}^{n-1}\}_{r=1}^{\mathcal{M}_{u,t}})$:
            \setcounter{ALG@line}{0}
            \For{each neighbor location $r \in \mathcal{N}_{u, t}$}
            \State $e_{u,r,t}^n = \left( \mathcal{N}_{r,t} e_{r,t}^{n-1} -  e_{u,t}^{n-1} + e_{u,t}^{n} \right)/ \mathcal{N}_{r,t}$
            \EndFor
            \State pseudo locations $\leftarrow$ \textbf{PseudoLocGenerate}(actual locations, $n_p$) 
            \For{each actual location and each pseudo location}
            \State $e_{u,r,t}^n \leftarrow \textbf{LocEmbeddingPerturb}(e_{u,r,t}^n, \sigma_l)$ \Comment{Perturb before uploading to the server.}
            \EndFor \\
            \Return local location embeddings $\{e_{u, r,t}^n\}_{r=1}^{n_p \times \mathcal{M}_{u,t}}$ to the server
        \Statex \hspace{-17pt} \textbf{ClientUpdate}($\{e_{r,t}^n\}_{r=1}^{\mathcal{M}_{u,t}}$):
            \setcounter{ALG@line}{0}
            \For{each neighbor location $r \in \mathcal{N}_{u, t}$}
            \State Extract hidden state $e_{m; r,t}$ from the macroscopic model
            \State $e_{r,t}^n \leftarrow \left[ e_{r,t}^n \parallel e_{m;r,t} \right]$ \hfill \Comment{Macroscopic model incorporation.}
            \EndFor
            \State ${\boldsymbol{x}}_u \leftarrow \frac{1}{T}\sum_{t=1}^T\frac{1}{\mathcal{M}_{u,t}}\sum_{r=1}^{\mathcal{M}_{u,t}} e_{r,t}^n$ \Comment{Aggregate from hyperedges to node.}
            \State $\hat{{y}}_u \leftarrow {\rm Predictor} \left(\boldsymbol{x}_u, \mathbf{\Theta}^{n-1}\right)$ \Comment{Predict the infection state.}
            \State  $\ell_u \leftarrow \ell\left(\hat{{y}}_u, {y}\right)$
            \State $e_{u}^{n} \leftarrow e_{u}^{n-1} - \eta_{lr} \nabla_{e_{u}^{n-1}} \ell_u$ \Comment{Update the client embedding.}
            \State $ \mathbf{g}_u \leftarrow \nabla_{\mathbf{\Theta}^{n-1}} \ell_u$ \Comment{Calculate the gradient of model parameters.}
            \State $\mathbf{g}_u=\operatorname{clip}\left(\mathbf{g}_u, C_f\right)+\mathcal{N}(0, \sigma_f^2)$  \Comment{Add DP noise to the gradient.}
           \\ \Return $\mathbf{g}_u$ to the server
        \Statex \hspace{-17pt}\textbf{PseudoLocGenerate}(actual locations, $n_p$)
        \setcounter{ALG@line}{0}
        \State $p_{r,t}^{r',t+1}(u) = p\left[l_{t+1}=r'|l_t=r\right] = \pi_u(r|t) \times p_u(r'|r)$ \Comment{Aggregate mobility model.}
        \State  $\overline{p}_{r,t}^{r',t+1} = \frac{1}{N}\sum_{u \in \mathcal{U}} p_{r,t}^{r',t+1}(u) + \epsilon \cdot max(1, d(r,r'))^{-2}$ \Comment{Aggregate global mobility model.}
        \State $sim(r,r') =  \sum_t^T \left(x_t-x'_t\right) + \gamma d(r,r')^{-2}$ \Comment{Epidemic Clustering.}
        \State Transform trace into epidemic domain
        \For{$i=1, 2, \cdots, n_p$}
        \State Sample trace from the epidemic domain via $\overline{p}_{r,t}^{r',t+1}$
        \EndFor
        \\ \Return pseudo locations
        \Statex \hspace{-17pt} \textbf{LocEmbeddingPerturb}($e_{u,r,t}^n$, $\sigma_l$):
            \setcounter{ALG@line}{0}
            \State $e_{u,r,t}^n \leftarrow e_{u,r,t}^n + \mathcal{N}(0, \sigma_l^2)$ \Comment{Add DP noise to the local location embedding.} \\
            \Return $e_{u,r,t}^n$
    \end{algorithmic}
    \end{algorithm}

\subsubsection{Detached Hypergraph Propagation}\label{par:detach}

\begin{figure}[t!]
    \centering
    \includegraphics[width=.9\textwidth]{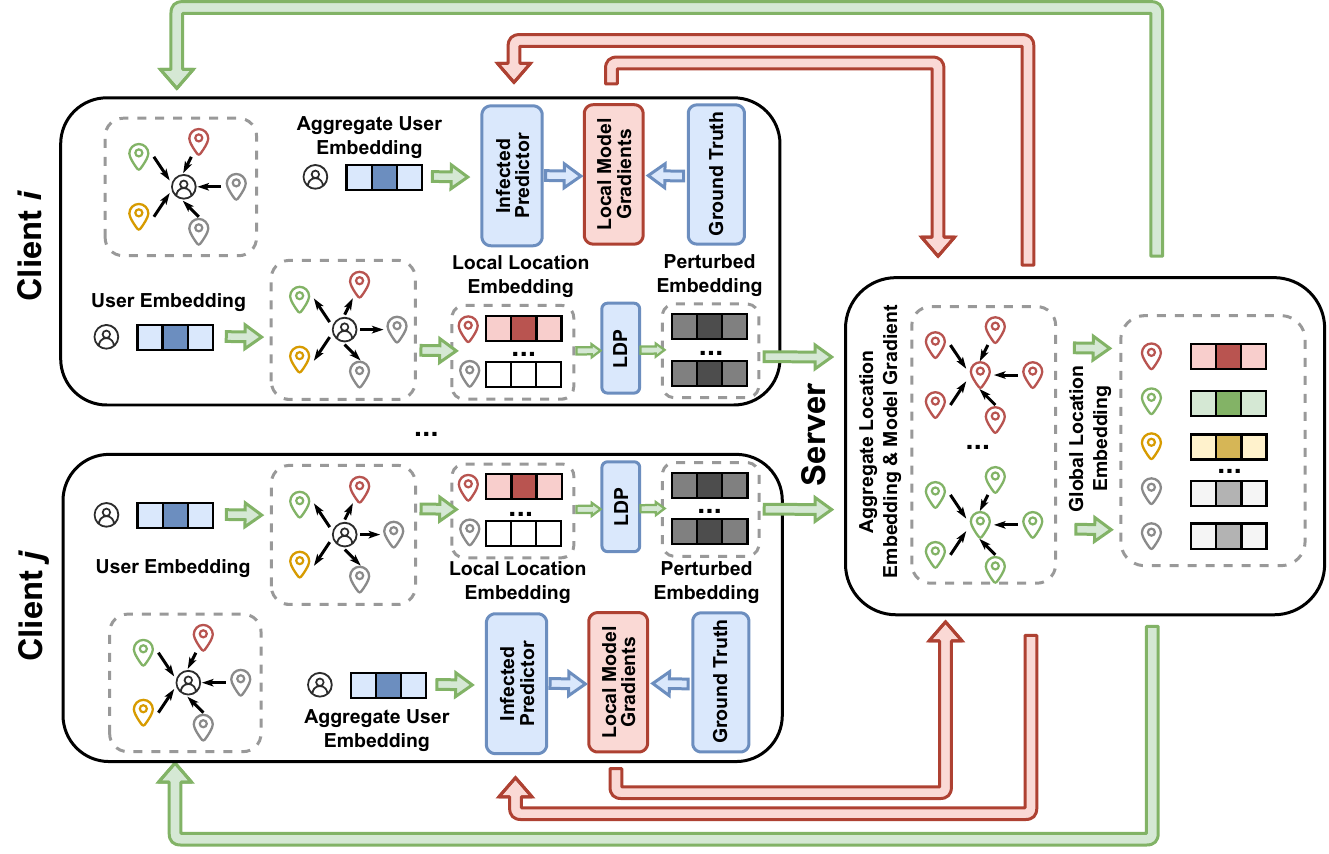}
    \caption{Illustration of privacy-preserving FGML framework.}
    \label{fig:fgml}
\end{figure}

In the scenario of FGML, an important challenge is that each client can only access a subgraph of the global graph and some nodes may be neighbors of other clients~\cite{fu2022federated}. However, a client can mostly only aggregate the features of nodes within the local storage under the constraint of privacy requirements. 
In the task of individual-level infection prediction, this challenge still exists and is even more intractable. Specifically, restricted by privacy considerations, each individual stores its trajectory data and features in the local device and conceals them from being accessed by other individuals. Thus, it is unattainable to construct the individual-individual contact graph in each client, which makes it more intractable for an individual to aggregate the features of other individuals.
To address this complex issue, we first opt for the spatio-temporal hypergraph to establish interaction information between individuals and locations, allowing each client to construct a subgraph from their local historical trajectory. 
In addition, as depicted in Figure~\ref{fig:fgml}, we decouple the two phases of hypergraph aggregation and use the hyperedge as the mediator to propagate individual information, allowing clients to share information by uploading embeddings of visited locations. Throughout the entire process, the historical trajectories of all users will not be shared with other users, and no user will be able to detect other users who have visited the same location.

Intuitively, the propagation from node to edge is deployed on the server side, which can be expressed as:
\begin{equation}\label{equ:detach1}
\mathbf{E}^{(l+1)}=  \mathbf{W} \mathbf{D}_e^{-1} \mathbf{H}^\top \mathbf{X}^{(l)},
\end{equation}
which represents the first part of (\ref{equ:HGNN}), and $\mathbf{E}^{(l+1)}$ denotes the aggregate hyperedge embedding, $\mathbf{X}^{(l)}$ represents the node embedding in this work. The hyperedge weight matrix $\mathbf{W}$ is set to an identity matrix in this work. The propagation from edge to node is executed by local clients, which can be represented as follows:
\begin{equation}\label{equ:detach2}
\mathbf{X}^{(l+1)}= \sigma 
\left( 
\mathbf{D}_v^{-1} \mathbf{H} \mathbf{E}^{(l+1)} \mathbf{\Theta}^{(l)}
\right),
\end{equation}
which represents the second part of (\ref{equ:HGNN}).

Consequently, \emph{Falcon} maintains the cross-client transition of information and prevents the exposure of the local subgraph structure to other clients. Nevertheless, the server still can infer where the user has visited with the inference attacks, which can estimate the real location of users via non-zero gradients~\cite{melis2019exploiting} or sequential location queries~\cite{shokri2011quantifying}. Thus, we propose a novel plausible pseudo location generation technique and introduce a differential privacy (DP) mechanism to guarantee that even for the central server, it can not directly infer the real location the user visited. 

\subsubsection{Differential Privacy Perturbation Mechanism}\label{par:dp-per-mec}
In the following content, we introduce a differential privacy perturbation mechanism, which is integrated into the aforementioned detached propagation mechanism to confront the inference attack via non-zero gradients~\cite{melis2019exploiting}.
Note that each client only stores user embedding on the local device, and according to (\ref{equ:detach1}), the server needs to aggregate user embeddings to location embeddings. Consequently, instead of simply downloading user embeddings, each client calculates the local embedding for each location beforehand. The detailed calculation of the local location embedding in the $n$-th epoch is as follows:

\begin{equation}
    e_{u,r,t}^n = \left( 
   \mathcal{N}_{r,t} e_{r,t}^{n-1} -  e_{u,t}^{n-1} + e_{u,t}^{n}
    \right)/ \mathcal{N}_{r,t} ,
\end{equation}
 where $e_{u,r,t}^n$ denotes the local embedding of location $r$ in client $u$ at time interval $t$, $e_{r,t}^{n-1}$ denotes the global location embedding download from the server. $e_{u,t}^{n}$ represents the user embedding stored in the client. $\mathcal{N}_{r, t}$ is the total number of individuals located in the region $r$ at time $t$. Thus, client $i$ will contribute embedding to each location $r$ in the history trace record: $L_u$, for the location not existing in $L_u$ (i.e., the pseudo location), the local location embedding equal to $0$. To avoid the honest-but-curious server inferring the real location by detecting the location embedding with the non-zero gradient, we propose a differential private perturbation mechanism satisfying the $(\epsilon,\delta)$-DP~\cite{wei2020federated}. Specifically, the updated value of a user for each location is clipped with a constant value $C_l$, and then Gaussian noise $\mathcal{N}(0, \sigma^2_l)$ is added to each location embedding. The standard deviation $\sigma_l$ can be calculated as follows:
 \begin{equation}
\sigma_{l}=\frac{L C_l \sqrt{2 \ln (1.25 / \delta)}}{\epsilon},
\end{equation}
where $L$ denotes the exposure number of parameters, that is, the number of training epochs. 
After executing the above processes, each user uploads their update of the corresponding location embedding to the central server. 

On the server side, the server aggregates all local location embeddings received from users and takes an average to obtain the aggregated location embedding, which can be described as follows:
\begin{equation}
    e_{r,t}^n = 
    \left(
      \sum_{u \in \mathcal{N}_{r,t}} e_{u,r,t}^n - (\mathcal{N}_{r,t} - 1) e_{r,t}^{n-1}
    \right) ,
\end{equation}
where $e_{r,t}^n$ is the aggregated location embedding in the $n$-th epoch. Then, each aggregated location embedding is distributed to all clients, and each client propagates the location embedding to node embedding following the (\ref{equ:detach2}).



\begin{figure}[t!]
    \centering
    \includegraphics[width=.8\textwidth]{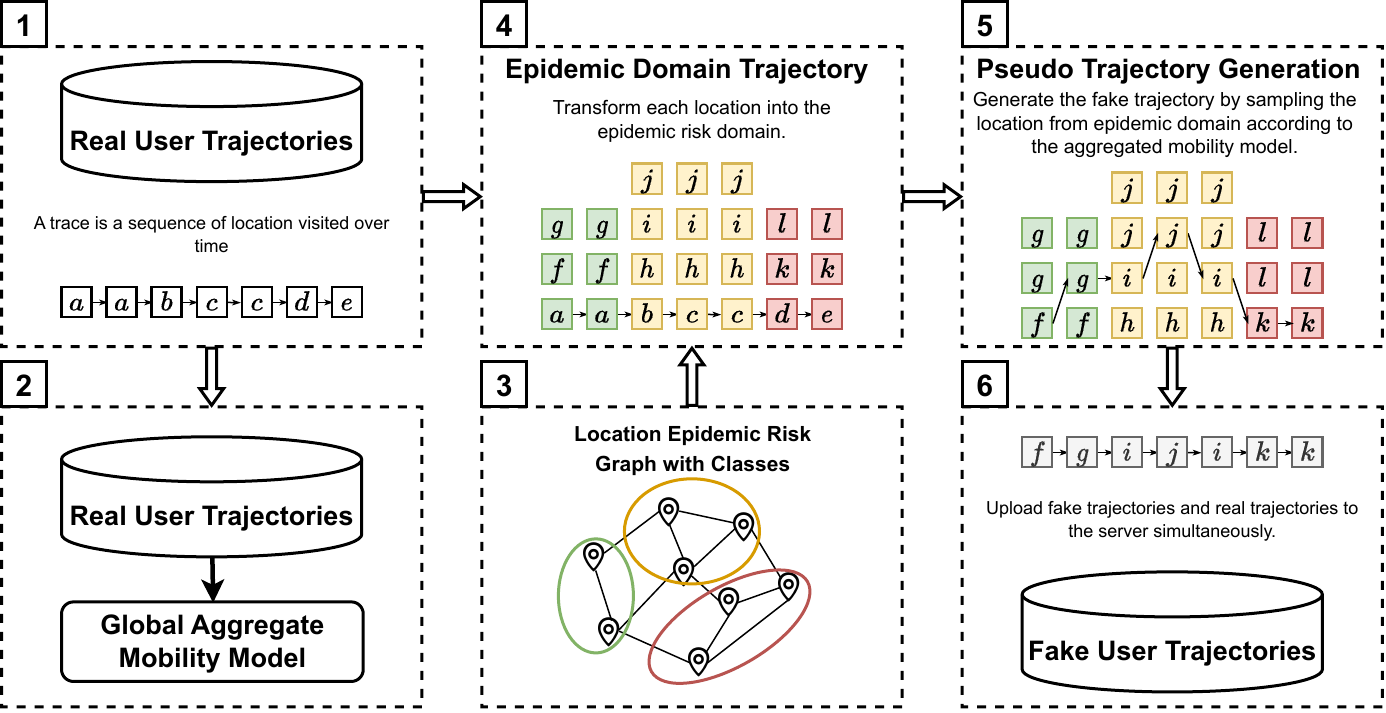}
    \caption{Illustration of plausible pseudo location generation.}
    \label{fig:pseudo_location}
\end{figure}

\subsubsection{Plausible Pseudo Location Generation}\label{par:pseudo}
Stipulating clients to upload embedding of the actual visited location $a_u(t)=r$ while also attaching several embeddings of randomly chosen pseudo locations $p_u(t)=(r'_1, r'_2, \cdots)$, and adding noise to all the embeddings. It can effectively prevent inference attacks through the non-zero gradient, and conceal the observation of the user's locations on the server side to be $o_u(t) = \left\{r, r'_1, r'_2, \cdots\right\}$. However, it cannot defend against location inference attacks via sequential location queries.
These attacks are able to filter out the genuine location over multiple location queries over time, if the pseudo location is not believable.
Thus, a more plausible strategy for generating pseudo locations should be investigated. We proposed a plausible pseudo location generation algorithm inspired by a synthesizing traces technique~\cite{Bindschaedler2016Synthesizing} as shown in Figure~\ref{fig:pseudo_location}, which can generate synthetic traces as opposed to independent pseudo locations. The details of this algorithm can be divided into the following four parts:


     \para{Aggregate Mobility Model:} Conceptually, as shown in Figure~\ref{fig:pseudo_location}.2 we construct the global aggregate mobility model by averaging over the local mobility models from sampled user groups $\mathcal{U}$. First, the local mobility model of a specific user $u$ can be represented by a Markov model, which could be formulated as follows:
     \begin{equation}
         p_{r,t}^{r',t+1}(u) = p\left[l_{t+1}=r'|l_t=r\right] = \pi_u(r|t) \times p_u(r'|r),
     \end{equation}
     where $\pi(r|t)$ denotes the probability of user $u$ visits to location $r$ at time $t$, and $p(r'|r)$ is the transition probability that indicates the willingness of user $u$ move from location $r$ to location $r'$. Thus $p_{r,t}^{r',t+1}(u)$ represents for the estimated probability for user $u$ to move from location $r$ to $r'$ at time $t$. Consequently, the aggregate global mobility model can be characterized as follows:
     \begin{equation}
         \overline{p}_{r,t}^{r',t+1} = \frac{1}{N}\sum_{u \in \mathcal{U}} p_{r,t}^{r',t+1}(u) + \epsilon \cdot max(1, d(r,r'))^{-2},
     \end{equation}
     where $N$ is the total number of users for normalization, and $\epsilon$ is a small constant to avoid zero probability, $d(r,r')$ denotes the distance between $r$ and $r'$.
     
     \para{Epidemic Clustering:} Different from the trajectory synthesizing approach proposed in ~\cite{Bindschaedler2016Synthesizing}, which simultaneously takes both geographic and semantic features into consideration. In the COVID-19 infection prediction task, the semantic features mentioned before are not such sensitive. Since the semantic similarity is specifically proposed for LBS, which guarantees the synthetic location will not perturb the semantic profile of each user and affect the service quality of the LBS (e.g., location recommendation). Different from the LBS needs the identity of semantic similarity among users for a better personalization service, the epidemiology similarity of location should be considered in synthesizing pseudo trajectory to maintain the invariance of the epidemiology profile. Therefore, we proposed a method to estimate the epidemiology similarity between locations, then construct a location epidemiology graph, where the nodes are locations and the edge weights are the epidemiology similarity between nodes. The similarity between the two locations can be calculated as follows:
     \begin{equation}
         sim(r,r') =  \sum_t^T \left(x_t-x'_t\right) + \gamma d(r,r')^{-2},
     \end{equation}
    where $x_t$ and $x'_t$ denote number of new cases number at region $r$ and $r'$ in time interval $t$, respectively. $\gamma$ is a constant coefficient of the distance between $r$ and $r'$. 
    Then we run a clustering algorithm on the epidemiology graph to divide the location set into several clusters as illustrated in Figure~\ref{fig:pseudo_location}.3, within a specific cluster, locations share a similar epidemiology status.

     \para{Transforming Traces into Epidemic Domain:} As demonstrated in Figure \ref{fig:pseudo_location}.4, we transform each location in a trajectory into its corresponding epidemiological domain trajectory by simply substituting each location with its epidemiological equivalent. Clearly, the trajectory of the epidemiological domain comprises all candidate locations that exhibit a similar epidemiological condition to the genuine one.

     \para{Sampling Trace from the Epidemic Domain:} 
     As depicted in Figure~\ref{fig:pseudo_location}.5, we finally decode the trajectory of the epidemiology domain into the geographic domain, which has the same format as the original actual trajectory. With a time-varying Markov model, we undertake a random walk on the epidemiology domain trajectory.
     To initialize the generation procedure, we sample a location from the first candidate location set in the epidemiology domain trajectory as the initial location. Then, we randomly select the next location based on the aggregate mobility model (i.e., the transition probability) $\overline{p}_{r,t}^{r',t+1}$. Note that multiple plausible trajectories can be generated by using different initiatory locations or repeatedly the sampling process with the time-varying Markov model.

\subsubsection{Privacy-preserving model updating}
    In our framework, although the two phases of aggregation are divided to be processed in clients and the server, the training method of the model still follows the general setting of FL. That is, the parameters of the model are stored in the local client of each user, and the clients update the global model by uploading the updated gradient or parameters of the local model. Nevertheless, several existing researches have claimed that the private information of users can still be divulged via the uploaded gradients~\cite{melis2019exploiting} or parameters~\cite{hitaj2017deep, shokri2015privacy}. Adding random noise is a natural approach to prevent these values from leaking too much information about the user, and one prominent example is differential privacy (DP)~\cite{shokri2015privacy, abadi2016deep}.
    Thus, we utilize DPSGD~\cite{abadi2016deep}, the widely used method satisfying DP, to train this distributed model. DPSGD will truncate the value of local model gradients into $\left[ -C_f, C_f\right]$ range and add Gaussian noise to gradients before uploading them to the server:
    \begin{equation}
        \mathbf{g}_u=\operatorname{clip}\left(\mathbf{g}_u, C_f\right)+\mathcal{N}(0, \sigma_f^2),
    \end{equation}
where $\mathbf{g}_u$ denotes the model gradient of user $u$, $C_f$ represents the clip threshold, $\sigma_f$ is the volume of noise. Note that the DP mechanism introduced here is not identical to the differential privacy perturbation mechanism mentioned in Section~\ref{par:dp-per-mec}. DPSGD protects the model parameters under the FL framework, while the perturbation mechanism protects the node features under the FGML framework.

\subsection{Macroscopic Model Cooperation}
\begin{figure}[t!]
    \centering
    \includegraphics[width=.8\textwidth]{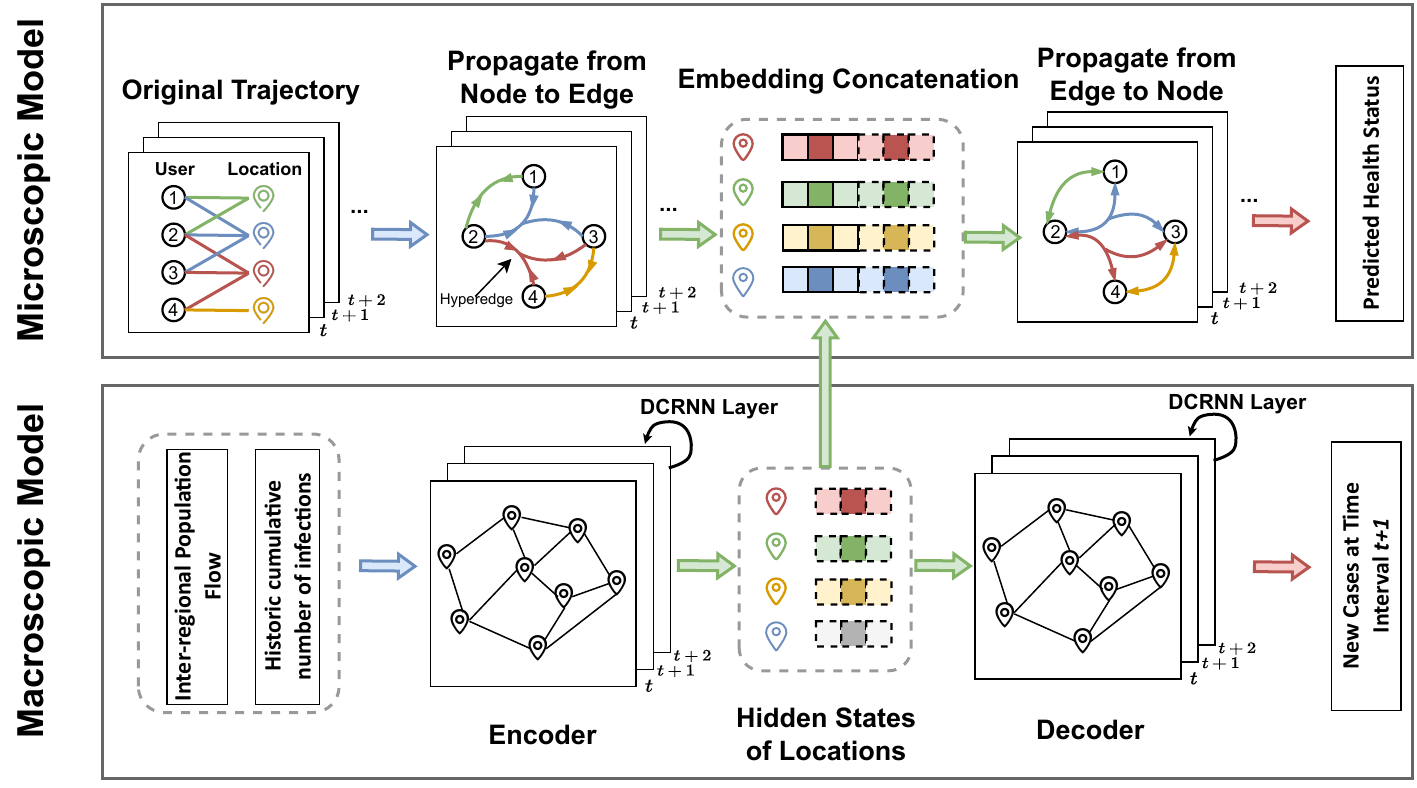}
    \caption{Illustration of the coupling mechanism between microscopic and macroscopic models.}
    \label{fig:macro_model}
\end{figure}

The privacy-serving FGML framework we proposed is capable of protecting the user from location privacy leakage and helping them share cross-client information simultaneously. However, the overall performance of the framework declines when we deploy the DPSGD for distributed model training, as well as the pseudo locations generation method and the coupled perturbation mechanism for hiding users' genuine locations. 
The perturbation mechanism and pseudo location generation method conceal the real locations of users by generating accompanying pseudo locations and adding noise to all location embeddings. These location obfuscation mechanisms diminish the information quality of spatio-temporal points, making it challenging for user nodes to aggregate accurate epidemic information from spatio-temporal points. Therefore, it is a natural approach to utilize an auxiliary model to re-extract the noise-free information of spatio-temporal points.
Specifically, as shown in Figure~\ref{fig:macro_model}, we construct a region-level infection prediction model as the auxiliary macroscopic model to recapture the epidemic information of location. Then, we design a novel cooperative coupling mechanism to integrate the microscopic model with the macroscopic model to counteract the performance decline.
Before diving into the details of the cooperative coupling mechanism, we first give a brief introduction to the macroscopic model.

\para{Macroscopic Model:} In contrast to the microscopic model that models infectious disease at the individual level, the macroscopic model is used to model the pandemic at the level of locations or regions. 
Specifically, the goal of the macroscopic model is to predict the number of new infection cases for $\mathcal{M}$ regions in the next $T$ time interval by using the inter-regional population flow and the historical number of new infection cases in each location.
In this paper, we utilize the spatio-temporal graph to model the population flow among different regions, then capture the epidemic diffusion with the GNN model.
The population flow networks among regions at time interval $t$ can be represented as a weighted directed graph $\mathcal{G}^{(t)}=\left(\mathcal{V},\mathcal{E},\mathbf{W}^{(t)}\right)$, where $\mathcal{V}$ is the set of regions, $\mathcal{E}$ is the set of the edges and $\mathbf{W}^{(t)} \in \mathbb{R}^{\mathcal{M} \times \mathcal{M}}$ is corresponding to the weight matrix of edges.
Let $\boldsymbol{X}^{(t)} \in \mathbb{R}^{\mathcal{M} \times 1}$ represent the new cases of each region observed in time interval $t$, the formal target of the macroscopic model is to fit a function $f$ that predicts the new cases in time $t+L$ with historical $T$ time interval records:



\begin{equation}
    \boldsymbol{X}^{(t+L)} = f\left(  \boldsymbol{X}^{(t-T-1)}, \cdots, \boldsymbol{X}^{(t)}  ;  \mathcal{G}^{(t-T-1)} , \cdots, \mathcal{G}^{(t)} \right).
\end{equation}
In this work, we apply DCRNN ~\cite{li2018diffusion}, a widely-used spatio-temporal graph neural network (STGNN), for this time series prediction task. In order to make DCRNN adapt to the dynamic changes of the inter-regional population flow, we revise its convolution layer to support the dynamic graph. Consequently, the refined convolution layer of DCRNN can be formulated as follows: 
\begin{equation}
\left\{\begin{array}{l}
\boldsymbol{r}^{(t)}=\sigma\left(\boldsymbol{\Theta}_r \star \mathcal{G}^{(t)}\left[\boldsymbol{X}^{(t)}, \boldsymbol{H}^{(t-1)}\right]+\boldsymbol{b}_r\right), \\[10pt]
\boldsymbol{u}^{(t)}=\sigma\left(\boldsymbol{\Theta}_u \star \mathcal{G}^{(t)}\left[\boldsymbol{X}^{(t)}, \boldsymbol{H}^{(t-1)}\right]+\boldsymbol{b}_u\right), \\[10pt]
\boldsymbol{C}^{(t)}=\tanh \left(\boldsymbol{\Theta}_C \star \mathcal{G}^{(t)}\left[\boldsymbol{X}^{(t)},\left(\boldsymbol{r}^{(t)} \odot \boldsymbol{H}^{(t-1)}\right)\right]+\boldsymbol{b}_c\right), \\[10pt]
\boldsymbol{H}^{(t)}=\boldsymbol{u}^{(t)} \odot \boldsymbol{H}^{(t-1)}+\left(1-\boldsymbol{u}^{(t)}\right) \odot \boldsymbol{C}^{(t)},
\end{array}\right.
\end{equation}
where $\boldsymbol{X}^{(t)}$, $\boldsymbol{H}^{(t)}$ denote the input and output hidden state at time interval $t$, $\boldsymbol{r}^{(t)}$, $\boldsymbol{u}^{(t)}$ are reset gate and update gate at time $t$ respectively. $\star \mathcal{G}^{(t)}$ denotes the diffusion convolution proposed in ~\cite{li2018diffusion}, which models the dynamics of the epidemic as a diffusion process. $\boldsymbol{\Theta}_r$, $\boldsymbol{\Theta}_u$, $\boldsymbol{\Theta}_C$ are parameters for corresponding filters.
Note that the DCRNN model can be easily substituted with other STGNN models like STGCN~\cite{yu2018spatiotemporal},  GraphWaveNet~\cite{wu2019graph} and D2STGCN~\cite{shao2022decoupled}.

\para{Cooperative Coupling Mechanism:}
The microscopic and macroscopic models we have proposed are still two independent frameworks. To incorporate the individual-level infection prediction model with the auxiliary region-level infection prediction model, we design a cooperative coupling mechanism that establishes a connection between the two models. The performance degradation of the microscopic model is largely blamed on excessive disturbance of spatio-temporal point embeddings during feature aggregation. Thus, we extract the output hidden state $\mathbf{E}_m$ from the encoder module of the macroscopic model, which includes noise-free information of spatio-temporal points and can serve as a powerful supplement in the microscopic model. 
Consequently, different from directly downloading hyperedge embeddings (i.e., the aggregate location embeddings) for aggregation, each client also obtains the hidden state from the macroscopic model and concatenates it to hyperedge embeddings. In this way, the propagation from edge to node deployed on the client side can be amended as follows:
\begin{equation}
\mathbf{X}^{(l+1)}= \sigma 
\left( 
\mathbf{D}_v^{-1} \mathbf{H} \left[ \mathbf{E}^{(l+1)} || \mathbf{E}_m \right] \mathbf{\Theta}^{(l)}
\right),
\end{equation}
where $||$ denotes the concatenation operation, $\mathbf{E}_m$ is the output hidden state of the encoder module in the macroscopic model. 
Notably, the number of new infections in each region is estimated based on the confirmed cases observed in each region, and the inter-regional population flux is determined by statistically analyzing individual trajectory data. In contrast to the microscopic model, the macroscopic model requires no additional data. In addition, information is propagated unidirectionally from the macroscopic model to the microscopic model. Thus, there are no information leakage channels in the microscopic model regarding the coupling component.

\subsection{Prediction and Optimization}\label{optimization}
We utilize a fully connected layer to convert the node embedding $\mathbf{X}^{(L)} \in \mathbb{R}^{\mathcal{N}  \times  F}$ from the last HGNN layer into the probability output $\mathbf{Y} \in \mathbb{R}^{\mathcal{N}  \times C}$:
\begin{equation}
    \label{equ:output}
    \mathbf{Y} = {\rm Softmax}\left(\mathbf{X}^{(L)}W^{F \times C}\right),
\end{equation}
where $X^{(L)}$ is the output of the last of HGNN layer, $W^{F \times C}$ is the weight matrix of the linear layer. ${\rm Softmax}$ function is used to transform each prediction value into probability.

Then, we utilize the optimizer to minimize the cross-entropy loss, which is evaluated based on the predicted values and the ground truth:
\begin{equation}
\mathcal{L} = \sum\limits_{n=1}^{\mathcal{N}}{\left(-\log \left( \frac{\exp(x_n[c_n])}{\sum_{c=1}^{C} \exp(x_n[c])} \right)\right)},
\end{equation}
where $x_n$ denotes the probability vector of node $n$, $c_n$ represents the target class of node $n$. Thus, $x_n[c_n]$ indicates the predictive value of the target class, and $x_n[c]$ indicates the predictive value of class $c$.

\subsection{Algorithm Analysis}
\subsubsection{Privacy Guarantee}\label{par:privacy guarantee}
In this section, we will systematically introduce the privacy guaranteed by our framework from two different perspectives, the client side and server side, which address the challenges of cross-client information missing and graph structure leakage, respectively.
\begin{itemize}[leftmargin=*]
    \item \textbf{Cross-client side:}
    In this work, we design a detached propagation method on the proposed spatio-temporal hypergraph, which ensures the trajectory information is stored in the local client and theoretically guarantees privacy on the cross-client side. Specifically, clients only share the local calculated gradients of hyperedge embedding to the server, then download aggregated gradients from the server for the following prediction procedure. Thus, the privacy of the client is guaranteed with respect to other clients.
    \item \textbf{Server side:}
    However, even with our proposed detached propagation mechanism, the central server can still infer the actual location using adversary algorithms such as inference attack via non-zero gradient ~\cite{melis2019exploiting} or location inference attack ~\cite{shokri2011quantifying}. Thus, we propose a perturbation mechanism and a pseudo location synthetic method to address these two adversary concerns:
    \begin{itemize}
        \item \textbf{Pseudo location synthesis method against location inference attack (Localization Attack ~\cite{shokri2011quantifying})}
        In this work, we proposed a pseudo location generation algorithm to protect the user's location privacy, i.e., hiding the user's actual location from the server and also preventing the inference of the full trajectory. The user's device sends a number of pseudo locations along with the real location. For example, if three pseudo locations are used, then the device will send four locations (one real and three pseudos) to the server. Our proposed method can ensure that the server does not know which location is real and it is not able to estimate the real location over multiple queries over time. Additional experiments are conducted in Section ~\ref{par:privacy&performance} to verify that our pseudo location synthesis method can defend against the widely-used location inference attack ~\cite{shokri2011quantifying}. The results demonstrate that our method can achieve a better balance between privacy and utility.
        \item \textbf{Perturbation mechanism against inference attack via gradient ~\cite{melis2019exploiting}:}
        During training, the gradient of embedding in a client is sparse with respect to the history visited locations. Besides, the client needs to upload the gradients of local location embedding to the server for aggregation. Thus, we proposed a perturbation mechanism that prevents the server from being able to infer the actual location with non-zero gradients. 
    \end{itemize}
\end{itemize}
\subsubsection{Computational Complexity}
Existing GNN-based models only model the interaction among users and can not characterize the complicated interactions between individuals and locations well. Besides, the user-user interaction graph will be too huge to calculate on it ($O(\mathcal{N}^2)$ time complexity, $\mathcal{N}$ denotes the number of individuals). We design a spatio-temporal hypergraph construction method as presented on the left of Figure ~\ref{fig:overview}. Regards each individual as a node and connects all nodes that visit a location in the same time duration to a hyperedge. This design not only captures the temporal and spatial dependencies simultaneously but also models the complex interactions between users and locations. Therefore, our model (without macroscopic model enhancement) can better prediction performance. At the same time, our method can greatly alleviate the computational complexity of GNN with $O(\mathcal{N} \times \mathcal{M})$ time complexity ($\mathcal{M}$ denotes the number of locations, which is usually much lower than the population number).

%% file: 5.evaluation.tex
\section{Experiments}\label{sec::exp}
\subsection{Experiment Settings}
\subsubsection{Simulation Environment}
In this paper, we build an epidemic prediction benchmark environment based on the simulator\footnote{Simulator: \url{https://hzw77-demo.readthedocs.io/en/round2/simulator_modeling.html}} provided in the Prescriptive Analytics for the Physical World (PAPW) Challenge\footnote{PAPW 2020: \url{https://prescriptive-analytics.github.io/}}, which has been widely used by existing researches~\cite{feng2022contact, feng2022precise, dong2020hierarchical}. This simulator provides an accurate approach to synthesizing human mobility, and then simulates disease transmission based on the mobility data. To reconstruct the realistic scenario more accurately, we substitute the synthetic mobility data with two real-world, fine-grained mobility datasets that indicate two distinct scenarios. Besides, we also provide two terms of disease transmission parameters for SARS-Cov-2 and Omicron variant, where the scenario of the Omicron variant is utilized as an extensive experiment to evaluate the generalization of \emph{Falcon} in Section~\ref{par:generalization}.

\subsubsection{Experimental Scenarios.} 
To conduct multi-scenario experiments, we collect and prepare two datasets of location information for two distinct scenarios, \textbf{Basic} and \textbf{Larger}. The information of the two scenarios is depicted in Table~\ref{tab:scenario}.
For the \textbf{Basic} scenario, we consider the diffusion of infectious disease within $\mathcal{M}$ regions in a district. Each region is considered as a POI (Point of Interest, e.g., school, restaurant, bank). This scenario corresponds to the situation that trajectory data is collected by fine-grained localization technologies such as GPS~\cite{bajaj2002gps}, Bluetooth~\cite{faragher2015location}, and Wi-Fi~\cite{yang2008estimating}, etc. 
For the \textbf{Larger} scenario, we consider the disease spread within a city of $M$ regions, where each area is a community with an average radius of approximately $0.79$km. This scenario corresponds to the situation that trajectories collected by coarse-grained localization technologies such as Cell-ID~\cite{trevisani2004cell}.

The aforementioned datasets of two scenarios were supplied by a well-known location-based services (LBS) provider. When users use location-based services, their coordinates will be uploaded to the cloud server. For the ethical use of this data, we ensure that every user is aware of and consents to collect their location information, and all user IDs are anonymized to prevent them from being identifiable. Besides, our affiliation has issued the IRB approval and determined this study as nonhuman subject research.

\begin{table}[ht]
    \centering
    \caption{The information of the datasets of Basic and Larger scenarios.}
    \resizebox{.8\textwidth}{!}{
    \begin{tabular}{c||c|c|c|c|c} 
    \hlineB{3}
         Scenario & Population numbers & Region numbers & Scope & Region type & Duration   \\ \hline
        Basic & 15279 & 11234 & Changping District & POI & 40 days \\ \hline
        Larger & 15738 & 3127 & Beijing City & Area & 14 days \\ 
        \hlineB{3}
    \end{tabular}
    }
    \label{tab:scenario}
\end{table}

\subsubsection{Simulator Settings.} 
In this research, the epidemic dynamic is characterized by an SEIR model~\cite{yang2020modified}, and the parameters are calibrated to align with $R_0$ of SARS-Cov-2 (5.7) provided by recent research~\cite {sanche2020high}.
To further evaluate the generalization ability of models as we will discuss in Section~\ref{par:generalization}, we further recalibrate epidemic parameters to align with $R_0$ of the Omicron variant (10.78) to implement an additional experiment.
Detailed information on the two terms of disease transmission parameters are presented in Table~\ref{tab:data_parameter}, where $\beta$ is the transmission rate of infected individuals, $1/\mu$ denotes the recovery period, $1/\alpha$ is the average duration of the latent period. $R_0=\beta/\mu$ represents the basic reproductive number, which indicates the average number of secondary cases attributable to infection by an index case after the case is recovered~\cite{sanche2020high}.
The number of epidemic simulation days $T$ is set to align with the trajectory length for both Basic and Larger scenarios. To obtain a reasonable infected ratio in SARS-Cov-2 and Omicron on the last day, we initialize 300 and 30 infected individuals for them, respectively. 

\begin{table}[ht]
    \centering
    \caption{The values and corresponding information sources of two terms of disease transmission parameters.}
    \resizebox{.8\textwidth}{!}{
    \begin{tabular}{c|c|c|c|c|c|c} 
    \hlineB{3}
         Setting & $\beta$ & $\alpha $ & $\mu$ & $R_0$ & Extinction Recall & Related Literature  \\ \hline
        SARS-Cov-2 & 0.405 & 0.2564 & \multirow{2}{*}{0.071} & 5.7 & 82.46\%  & ~\cite{ghosh2022study, sanche2020high,zhao2021estimating}\\ \cline{1-3} \cline{5-7}
        Omicron variant & 0.766 & 0.6579 & & 10.78 & 90.72\%  &\cite{ghosh2022study, cai2022modeling}\\ \hlineB{3} 
    \end{tabular}
    }
    \label{tab:data_parameter}
\end{table}

\subsubsection{Evaluation Protocol}
In our experiments, we assume that we have prior knowledge of 40\% positive cases, and try to track the remaining potential infected individuals.
To comprehensively measure the performance of various baseline methods, we evaluate these methods with multiple metrics, including AUC, F1-score, Accuracy, and BEP (Break-Even Point, indicating the value when recall and precision are equal). Please note that the F1-score and accuracy we will release later are both the maximum values the model can achieve in the trade-off between precision and recall. To propose a more intuitive metric for the infection prediction task, we introduce a metric called disease extinction precision (DEP). 
The DEP assesses the maximum prediction precision that an individual-level infection prediction model can achieve while maintaining the minimum recall rate.
In epidemiological theory, the total number of infections will increase by a factor of $R_0$ after each average recovery period $1/\mu$.
Thus, we need to detect and control a ratio of $(R_0-1)/R_0$ positive cases to guarantee the extinction of disease, which corresponds to the minimum required recall: $r_m = 1 - 1/R_0$.
Based on the minimum recall for disease extinction, we can give the definition of DEP:
\begin{definition}[Disease Extinction Precision]
DEP is the precision that the prediction model can offer while maintaining the minimal recall of $r_m$.
This indicates that we must maintain a recall higher than $r_m$ to ensure disease extinction. Consequently, the value of DEP is the precision while recall reaches $r_m$.
\begin{equation}
        DEP = pr \left(r_m\right), 
\end{equation}
where $pr(\cdot)$ represents the function corresponding to the precision-recall curve.
\end{definition}

\subsubsection{Baseline Methods}
To conduct a comprehensive comparison, we selected a sequence of traditional methods and recent state-of-the-art works, resulting in a total of twelve baseline algorithms being employed in our experiments. Comprises an vanilla GCN model~\cite{kipf2017semi}, two attention-based GCN models~\cite{velivckovic2017graph, shi2020masked}, an infection prediction GNN model~\cite{tomy2022estimating}, five STGNN models~\cite{yu2018spatiotemporal, li2018diffusion, wu2019graph, shao2022decoupled, lan2022dstagnn}, one FGML framework~\cite{wu2022federated}, and a HGNN model~\cite{feng2019hypergraph}. Furthermore, DCT is also adopted as a baseline method for estimating the utility of existing digital contact tracing techniques~\cite{munzert2021tracking}. Out of all baseline methods mentioned above, only the FGML framework - FedPreGNN~\cite{wu2022federated} takes user privacy into consideration, which still fails to address the privacy divulgence of graph structures~\cite{fu2022federated}. 
\begin{itemize}[leftmargin=*]
    \item \textbf{DCT~\cite{munzert2021tracking}:} Digital contact tracing utilizes modern communication technologies, such as GPS, Wi-Fi, and Bluetooth, to digitally track and alert users who have interacted with infected individuals~\cite{munzert2021tracking}. In this work, DCT will track all potential infected individuals whose trajectories have intersected with those who have tested positive.
    \item \textbf{GCN~\cite{kipf2017semi}:} The basic semi-supervised classification algorithm on graph-structured data. It extends the convolutional operation into the graph structure by adopting ChebNet's first-order approximation in layer-wise propagation. 
    
    \item \textbf{GAT~\cite{velivckovic2017graph}:} A well-known revised version of the GCN model. GAT endows the model to focus on important links by assigning different weights to neighbor nodes with the attention mechanism. Moreover, GAT incorporates multi-head attention to stabilize the learning process.

    \item \textbf{UniMP~\cite{shi2020masked}:} A GCN-based semi-supervised classification algorithm that can simultaneously perform feature embedding and label embedding as input information for propagation within a graph transformer network. Besides, a masked label prediction strategy is proposed to avoid overfitting.

    \item     \textbf{EPIGNN~\cite{tomy2022estimating}:} A GCN-based individual-level infection prediction algorithm that is capable of estimating the infection state of all individuals in a large contact network.

    \item \textbf{STGCN~\cite{yu2018spatiotemporal}:} Spatio-Temporal Graph Convolutional Networks is a GCN model that captures spatial dependency and temporal dependency simultaneously by integrating graph convolution and gated temporal convolution through the spatio-temporal convolutional block. 

    \item \textbf{DCRNN~\cite{li2018diffusion}:} Diffusion Convolutional Recurrent Neural Network models the spatio-temporal dependency by proposing a novel Diffusion Convolutional Gated Recurrent Unit (DCGRU), which replaces the fully connected layer in GRU with a graph convolutional layer.

    \item \textbf{GraphWaveNet~\cite{wu2019graph}:} GraphWaveNet stacks the diffusion convolution layer and the dilated casual convolution layer [46] to jointly capture spatial-temporal dependency.

    \item \textbf{D2STGNN~\cite{shao2022decoupled}:} Decoupled Dynamic Spatial-Temporal Graph Neural Network proposes a Decoupled Spatial-Temporal Framework (DSTF) to separate the diffusion signal and inherent signal from the original data. 

    \item     \textbf{DSTAGNN~\cite{lan2022dstagnn}:} Dynamic Spatial-Temporal Aware Graph Neural Network introduces a novel dynamic spatial-temporal aware graph based on a data-driven strategy to substitute the static graph, as well as a multi-head attention mechanism to capture dynamic relations among nodes.

    \item \textbf{FedPreGNN~\cite{wu2022federated}:} FedPreGNN introduces a privacy-preserving graph expansion method to alleviate the information isolation problem under the federated learning scenario.

    \item \textbf{Hypergraph Neural Networks (HGNN)~\cite{feng2019hypergraph}:} The pioneering research extended the GCN model from a normal graph to a hypergraph, which can capture the high-order interaction among more than two nodes. 
    To adopt this model to the individual-level infection prediction task, we utilize the hyperedge of HGNN to represent a spatio-temporal point. Thus, it can be considered as an express edition of our framework without specific modules, such as FL, obfuscation mechanism, cooperative coupling mechanism, etc.
\end{itemize}

\subsubsection{Experiment Environment}\label{par:para_setting}
All experiments are compiled and tested on a Linux server (CPU: AMD EPYC-7763, GPU: NVIDIA GeForce RTX 3090). 
The default settings of hyperparameters are as follows. 
For model optimization, we set the learning rate to $0.001$, the dropout rate is set to $0.2$, the training epoch numbers are set to $500$, and the weight decay is set to $0.0005$. 
For the mobility and the epidemic data, the granularity of the time interval in this work is set to $2$ hours.
We assume that the infection status of 40\% of the total population is obtainable, meaning that 40\% of individuals are used as training samples. The Adam optimizer is utilized to train the prediction model.
For the perturbation mechanisms, the pseudo trajectories number is set to $2$ as default, and we set $\epsilon=1$, $\delta=0.001$, the clip threshold $C=0.1$ for the two differential privacy mechanisms (i.e. the DPSGD and the differential privacy perturbation mechanism).
In this work, our individual-level infection prediction model consists of an embedding layer for feature extraction, two HGNN layers for information propagation, and a fully connected layer for prediction.


\begin{table*}[t!]
\centering
\caption{Performance comparison of our model and baselines in two scenarios, where higher values represent better performance. \textbf{Bold} denotes the best results for each metric, and "-" denotes that the metric can not be evaluated for this baseline method.}
\label{tab:performance}
\resizebox{1\textwidth}{!}{
\begin{tabular}{c||c|c|c|c|c|c|c|c|c|c}
\hlineB{3}
\textbf{Scenarios} & \multicolumn{5}{c|}{\textbf{Basic Scenario}} & \multicolumn{5}{c}{\textbf{Larger Scenario}} \\ \hline
\textbf{Metrics}& DEP & AUC & F1 score & Accuracy & BEP & DEP & AUC & F1 score & Accuracy & BEP  \\ \hline \hline
\textbf{DCT} & - & - & 0.5052 & 0.7878 & -
& - & - & 0.3949 & 0.3311 & - \\ \hline
\textbf{GCN} & 0.3496 & 0.7957 & 0.5323 & 0.7535 & 0.5111
 & 0.2798 & 0.6666 & 0.4285 & 0.5758 & 0.3687   \\ \hline
\textbf{GAT} & 0.3440 & 0.7863 & 0.5262 & 0.7659 & 0.5106
& 0.2797 & 0.6679 & 0.4241 & 0.5647 & 0.3698   \\ \hline
\textbf{UniMP} & 0.2854 & 0.7204 & 0.4544 & 0.7096 & 0.4145
& 0.2552 & 0.6095 & 0.3929 & 0.5117 & 0.3127   \\ \hline 
\textbf{EPIGNN} & 0.3649 & 0.8103 & 0.5496 & 0.7803 & 0.5329
& 0.2943 & 0.6943 & 0.4474 & 0.6018 & 0.3774  \\ \hline 
\textbf{STGCN} & 0.3379 & 0.7691 & 0.4985 & 0.7104 & 0.4708
& 0.2621 & 0.6298 & 0.4054 & 0.4797 & 0.3354 \\ \hline
\textbf{DCRNN} & 0.3423 & 0.7851 & 0.5356 & 0.7606 & 0.4706
& 0.3144 & 0.7121 & 0.4700 & 0.6441 & 0.4067 \\ \hline
\textbf{GraphWaveNet} & 0.3565 & 0.8125 & 0.5570 & 0.7858 & 0.5439
& 0.3209 & 0.7254 & 0.4789 & 0.6609 & 0.4305   \\ \hline
\textbf{D2STGNN} & 0.3528 & 0.8025 & 0.5504 & 0.7841 & 0.5237
& 0.3245 & 0.7299 & 0.4838 & 0.6950 & 0.4314 \\ \hline
\textbf{DSTAGNN} & 0.3625 & 0.8161 & 0.5634 & 0.7932 & 0.5543
& 0.3220 & 0.7342 & 0.4907 & 0.7034 & 0.4492   \\ \hline 
\textbf{FedPreGNN} & 0.3434 & 0.7747 & 0.5078 & 0.7221 & 0.4691
& 0.2696 & 0.6342 & 0.4103 & 0.5500 & 0.3365 \\ \hline
\textbf{HGNN} & 0.4234 & 0.8397 & 0.6009 & 0.8132 & 0.5873
& 0.3389 & 0.7511 & 0.5016 & 0.7181 & 0.4653   \\ \hline
\textbf{Falcon(Ours)} & \textbf{0.5184} & \textbf{0.8945} & \textbf{0.6457} & \textbf{0.8306} & \textbf{0.6195}
& \textbf{0.4219} & \textbf{0.8257} & \textbf{0.5729} & \textbf{0.7527} & \textbf{0.5267}  \\ \hline
\textbf{Improvement} & \textbf{22.43\%} & \textbf{6.52\%} & \textbf{7.46\%} & \textbf{2.14\%} &  \textbf{5.48\%}
& \textbf{24.49\%} & \textbf{9.93\%} & \textbf{14.21\%} & \textbf{4.82\%} & \textbf{13.20\%}    \\ \hlineB{3}
\end{tabular}
}
\end{table*}

\begin{figure}[t!]
    \centering
    \includegraphics[width=.9\textwidth]{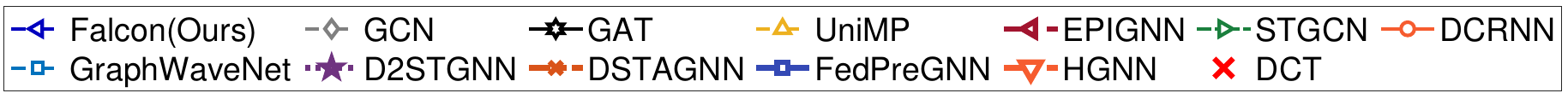}
    \subfigure[Basic Scenario]
    {\includegraphics[width=.4\textwidth]{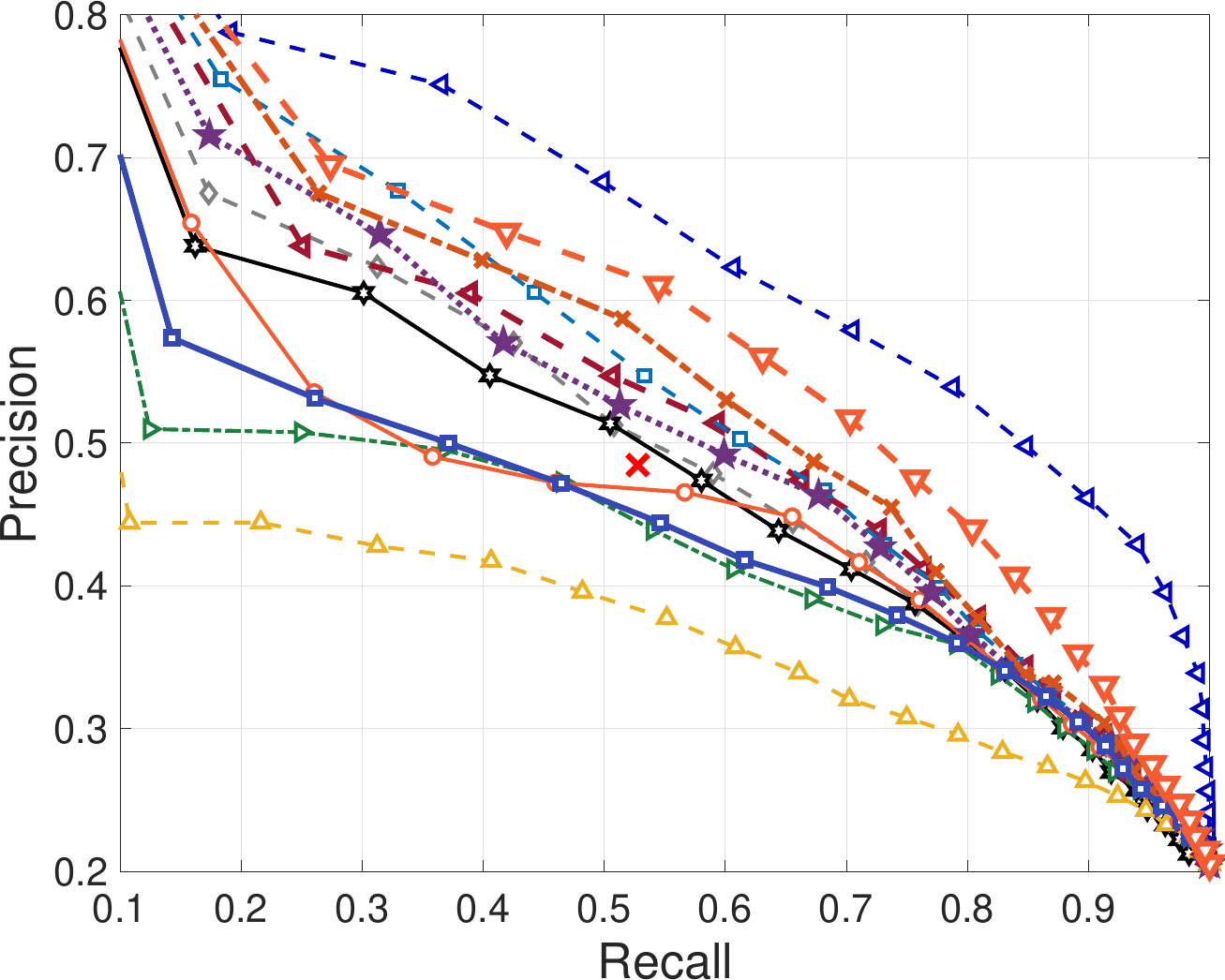}}
    \hspace{1.0cm}
    \subfigure[Larger Scenario]
    {\includegraphics[width=.4\textwidth]{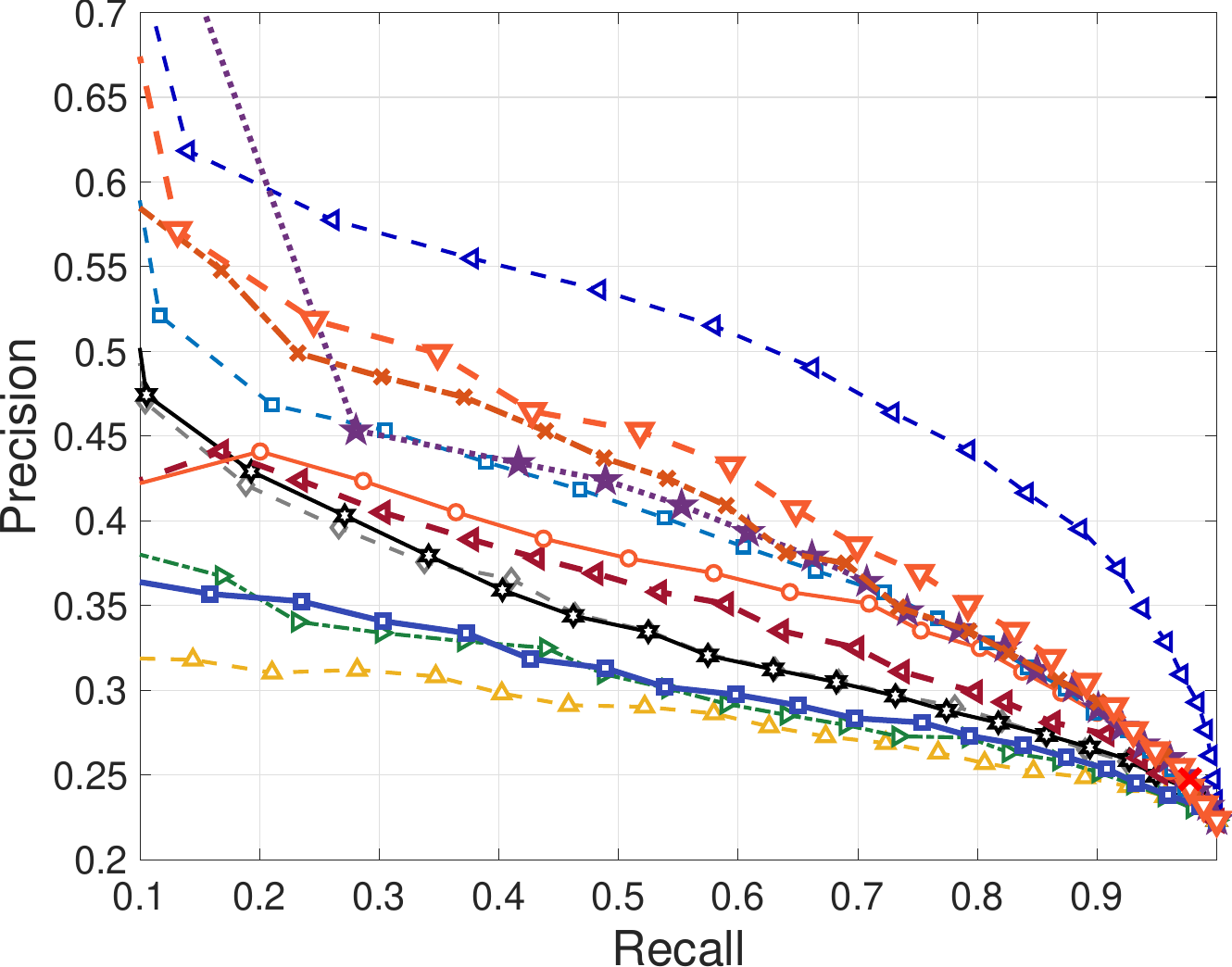}}
    \caption{The precision-recall (PR) trade-off curves of all baseline methods on the two scenarios.}\label{fig:trade-off}
\end{figure}

\subsection{Overall Performance Comparison}\label{par:performance comparision}
As shown in Table \ref{tab:performance}, we first compare \emph{Falcon} with baselines for all evaluation criteria under the Basic scenario and the Larger scenario, including multiple metrics for the classification task and our proposed metric DEP. In addition, we illustrate the Precision-Recall curves for each algorithm in Figure \ref{fig:trade-off} for a more comprehensible presentation. From the aforementioned experiments, the following analysis can be summarised:
\begin{itemize}[leftmargin=*]
    \item \para{Our proposed model significantly outperforms existing baseline methods in both Basic and Larger scenarios even within a strict privacy budget:} 
    Under the multi-scenarios, \emph{Falcon} always achieves the best infection prediction performance, although with several obfuscation mechanisms for location privacy-preserving. Particularly, with the spatio-temporal hypergraph construction and the cooperative coupling mechanism, \emph{Falcon} can improve the prediction precision by at least $22.43\%-24.29\%$ in multiple realistic scenarios. Besides, compared with the existing state-of-the-art FGML framework, i.e., FedPreGNN~\cite{wu2022federated}, our method remarkably improves the prediction precision by about $50.96\%-56.49\%$. Furthermore, as demonstrated in Figure~\ref{fig:trade-off}, our method always strikes the best balance between precision and recall.
    \item \para{DCT achieves fairly weak and inflexible performance in practical applications, especially with coarse-grained mobility data:} 
    The basic concept of DCT is to directly classify individuals who have had close contact with confirmed infected individuals to the high-risk population. Therefore, it is hard for DCT to provide a flexible trade-off between precision and recall, which is critical for various requirements of intervention strength. Specifically, as shown in Figure~\ref{fig:trade-off}, the recall achieved by the DCT in the Basic scenario is not sufficient to control the outbreak of the pandemic, and the extremely low accuracy achieved in the Larger scenario will lead to a waste of numerous medical resources. 
    In contrast, GNN-based models still strike relatively better prediction performance in both two scenarios, while simultaneously providing continuously adjustable intervention strength. It further verifies the necessity of designing a machine learning-enhanced DCT technology.
    \item \para{STGNNs generally strike better performance by capturing the spatio-temporal dependency:} Except for our proposed algorithm and HGNN that utilize the hyperedge to model the spatio-temporal point, STGNNs show the most promising performance among all baselines. For STGNNs, such as D2STGNN~\cite{shao2022decoupled} and DSTAGNN~\cite{lan2022dstagnn}, they both capture the spatial and temporal dependencies during disease transmission with GNNs and temporal units. Among them, DSTAGNN can achieve better performance than other STGNNs. The underlying reason for this phenomenon is that DSTAGNN effectively models the dynamic relationships among nodes, providing a more accurate representation of the evolving contact networks during the pandemic. The overall result shows it is crucial to model the spatio-temporal dependency in the infection prediction task.
    \item \para{The implementation of the attention mechanism on GNN led to similar or even worse predict performance:} 
    The attention enhancement GNN models, GAT and UniMP, neither of them provide performance gains for the prediction tasks. A potential explanation is that the attention can not assess correct weights to edges with limited node features in disease prediction situations. This phenomenon also points out the importance of estimating the weights of different interactions among individuals, which reveals the key to improving prediction precision for future research.

\end{itemize}

\subsection{Generalization Ability}\label{par:generalization}
In this section, we evaluate the generalization ability of \textit{Falcon} from the following two perspectives:

\begin{figure}[t!]
    \centering
    \includegraphics[width=.9\textwidth]{image/legend.pdf}
    \subfigure[Basic Scenario] 
    {\includegraphics[width=.4\textwidth]{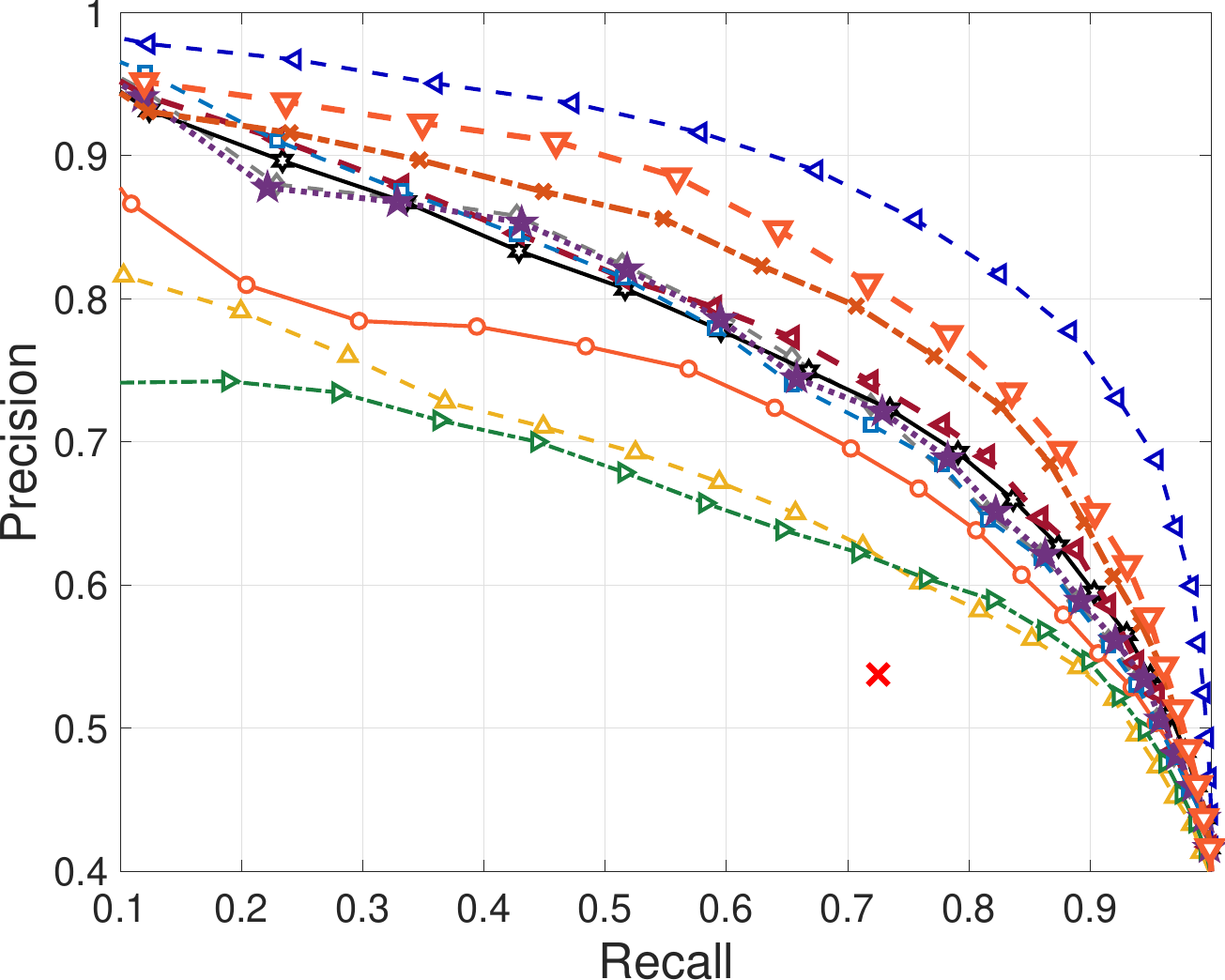}}
    \hspace{1.0cm}
    \subfigure[Larger Scenario]
    {\includegraphics[width=.4\textwidth]{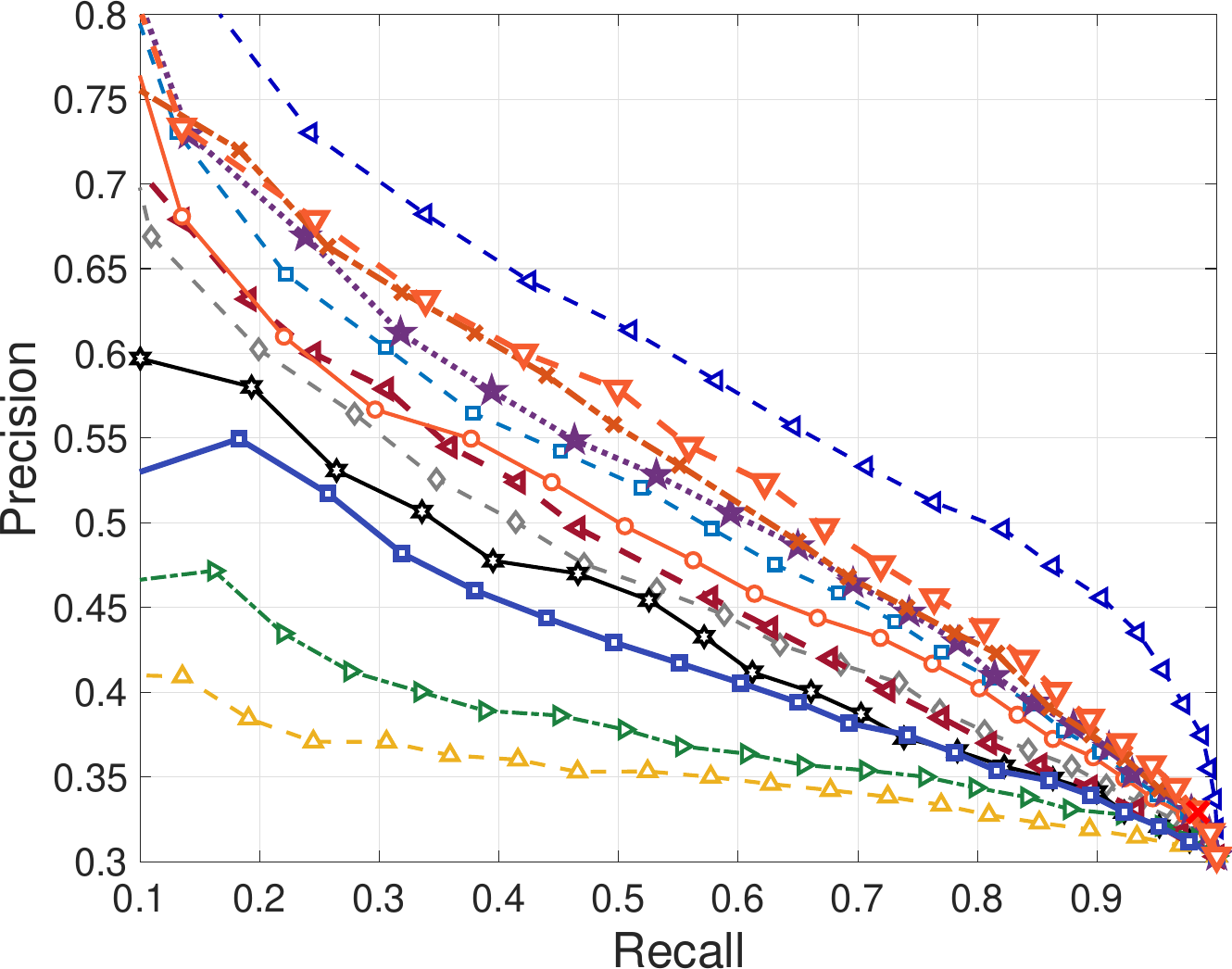}}
    \caption{The precision-recall (PR) trade-off curves of all baseline methods on the two scenarios with the Omicron settings.}\label{fig:trade-off-omicron}
\end{figure}

\para{Transmission Rates of Diseases:}
Different infectious diseases, and even different variants of the same infectious disease, have different transmission rates.
To verify the generalization ability of \emph{Falcon} against different COVID-19 strains with varying transmission rates, we implement extended experiments under the disease parameters of the Omicron variant. The results shown in Figure~\ref{fig:trade-off-omicron} demonstrate that under the Omicron setting, \emph{Falcon} can also strike the best performance, and maintain a better balance between precision and recall. Thus, \emph{Falcon} can effectively and efficiently track potential infected individuals for controlling the spread of disease.

\begin{figure}[t!]
    \centering
    \includegraphics[width=.7\textwidth]{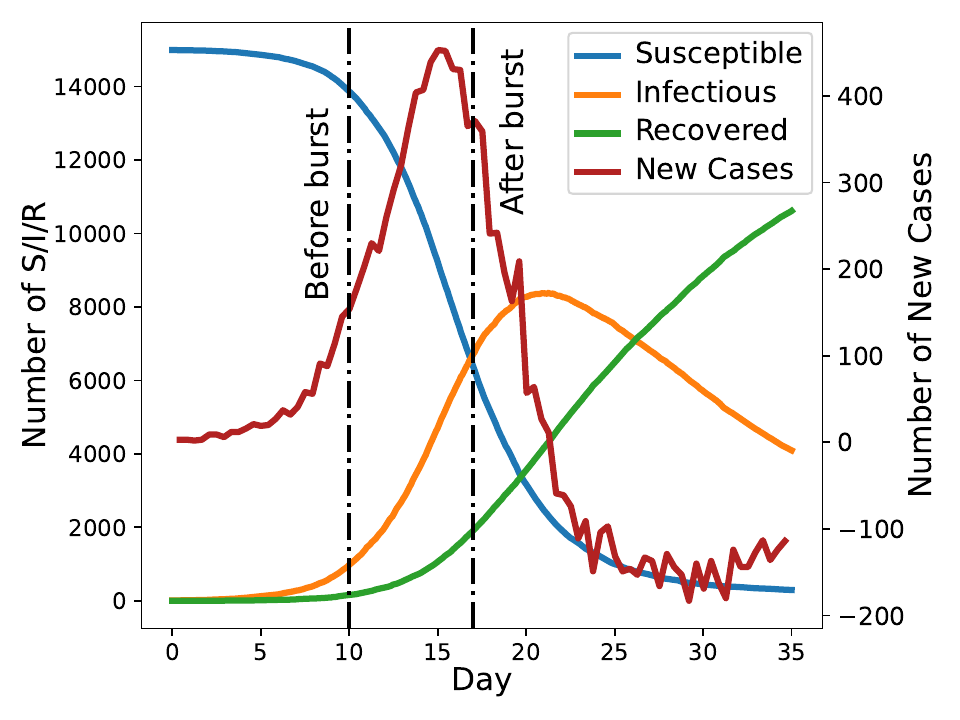}
    \caption{Evolution of the cumulative number of susceptible, infectious, and recovered individuals, as well as the daily number of newly confirmed infectious cases.}
    \label{fig:sir}
\end{figure}

\para{Stages of Pandemic:} 
In various stages of disease transmission, outbreaks can occur. Therefore, we investigate the prediction performance of \textit{Falcon} and representative baselines in different stages. The experiments are taken under the Large scenario and the Omicron settings. As shown in Figure~\ref{fig:sir}, we first plot the cumulative number of susceptible, infectious (including asymptomatic and symptomatic infectious.) and recovered individuals for each day, then provide the daily number of newly confirmed infectious cases. We notice that the growth rate of the number of infections reaches its peak around the 14th to 15th day, indicating the occurrence of an outbreak phenomenon. Therefore, we respectively evaluate each algorithm on the 10th and 14th day as two representative scenarios before and after the outbreak. The results are summarized in Table \ref{tab:burst}, which indicates that compared with other algorithms, \textit{Falcon} achieves better performance both before and after the outbreak. Moreover, during the early stages of the pandemic (before the outbreak), \textit{Falcon} is able to ensure a precision higher than 20\% for only a small percentage (6\%) of the infectious individuals. Therefore, at the same cost, \textit{Falcon} can detect more potential infected individuals, thus contributing to a faster control of the pandemic.

\begin{table}
\centering
\caption{Prediction performance of \textit{Falcon} and representative baselines on stages that before and after outbreak.}
\label{tab:burst}
\begin{tabular}{c||ccccccc}
\hlineB{3}
Phase        & \multicolumn{3}{c}{Before Outbreak}                          &  & \multicolumn{3}{c}{After Outbreak}  \\ 
\cline{2-4}\cline{6-8}
Method       & DEP               & F1                & AUC               &  & DEP    & F1     & AUC            \\ 
\hhline{=::=======}
DCT          & -                 & 0.1889            & -                 &  & -      & 0.7453 & -              \\
GCN          & 0.0975            & 0.3338            & 0.7527            &  & 0.6302 & 0.7495 & 0.6710         \\
GraphWaveNet & 0.1276            & 0.3457            & 0.7864            &  & 0.6375 & 0.7536 & 0.6931         \\
FedPreGNN    & 0.0932            & 0.3329            & 0.7358            &  & 0.5831 & 0.7249 & 0.6173         \\ 
\hline
Falcon       & $\mathbf{0.2086}$ & $\mathbf{0.4696}$ & $\mathbf{0.8952}$ &  & $\mathbf{0.6791}$ & $\mathbf{0.7753}$ & $\mathbf{0.7650}$  \\
\hlineB{3}
\end{tabular}
\end{table}



\subsection{Evaluation of Privacy and Utility}\label{par:privacy&performance}
As discussed in Section \ref{par:privacy guarantee}, there exist two privacy concerns with regard to the server side: the localization attack and the inference attack. We introduce a pseudo location synthesis method and a perturbation mechanism to address the two aforementioned privacy concerns. Therefore, we conduct experiments to assess the extent of the privacy budget provided by these two privacy-preserving methods and the amount of performance decline they incur.

\para{Perturbation mechanism against inference attack via gradient:} We perform a revised inference attack via gradient \cite{melis2019exploiting} on the honest-but-curious server. Specifically, the attacker sets a threshold for the 2-norm of the gradient. Values above this threshold are considered real locations, while those below are considered pseudo locations. The intensity of the proposed perturbation mechanism is depicted by the magnitude of the Gaussian noise. Thus, we summarize the evolution of the prediction performance and the attack error rate with regard to noise magnitudes, where we set up the number of pseudo locations as nine. As depicted in Table \ref{tab:performance noise}, as the magnitude of the noise increases, the attack error rate gradually rises and eventually reaches 90\% (random guess). When no noise is added, the error rate of the attack is 0, since the server can directly infer the actual location by detecting the location with non-zero gradients. Moreover, with the mitigation of the macroscopic model, the performance decline is relatively acceptable with the increasing intensity of the perturbation mechanism.

\begin{table}
\centering
\caption{The prediction performance and attack error rate w.r.t different noise magnitudes $\sigma_l$.}
\label{tab:performance noise}
\begin{tabular}{cccccc} 
\hlineB{3}
Noise magnitude $\sigma_l$ & 0      & 0.05   & 0.10   & 0.15   & 0.20    \\ 
\hline\hline
DEP                        & 0.4237 & 0.4233 & 0.4228 & 0.4203 & 0.4124  \\
AUC                        & 0.8270 & 0.8266 & 0.8261 & 0.8241 & 0.8212  \\
F1 score                   & 0.5746 & 0.5742 & 0.5736 & 0.5723 & 0.5708  \\ 
\hline
Attack error rate          & 0      & 0.773  & 0.877  & 0.892  & 0.896   \\
\hlineB{3}
\end{tabular}
\end{table}

\para{Pseudo location synthesis method against localization attack:} To measure how much privacy budget a plausible pseudo location generation algorithm can provide, we deploy a location inference attack \cite{shokri2011quantifying} on the server side. We consider the server of the LBS provider as an adversary, who is capable to captures the sequential location queries $O_u = \left\{o_u(1), o_u(2), \cdots,o_u(t) \right\}$ of a user $u$ and has the prior knowledge of the population mobility profile $\left<\overline{p}, \overline{\pi}\right>$. 
As we mentioned in Section~\ref{par:pseudo}, a user who queries the LBS at a time interval will upload a number of pseudo locations $p_u(t)$ along with its genuine location $a_u(t)$.
Correspondingly, a widely-adopted location inference attack - localization attack~\cite{shokri2011quantifying}, is deployed in the LBS server. Given historical observation (i.e., the sequence of locations queried to the LBS), the localization attack concentrates on discovering the user's true location at each time interval. The formal answer to such a goal is to estimate
\begin{equation}
    \operatorname{Pr}\left\{a_u(t)=r \mid O_{u}, \left<\overline{p}, \overline{\pi}\right>\right\}
\end{equation}
for each $r \in \mathcal{R}$. These probabilities can be easily computed with the Forward-Backward algorithm~\cite{rabiner1989tutorial}. Then, the adversary can form a distribution of possible regions, from which it can choose the most probable one.
We utilize the error probability of the inference attack as the privacy level of the pseudo location generation algorithm, where the Larger value means the higher privacy guarantee.
Different pseudo location selection mechanisms not only provides distinct privacy budgets but also curse the utility of the infection prediction model to various extent.
Therefore, we also evaluate the utility loss caused by the pseudo location generation technique. Here, we evaluate the performance decline with two metrics, AUC and F1-score. 

In this section, we compare our pseudo location generation algorithms with the following three existing pseudo location generation methods:
\begin{itemize}[leftmargin=*]
    \item \textbf{Uniform IID~\cite{shokri2011quantifying}:} we synthesize each pseudo location independently and identically from the uniform probability distribution. Therefore, the pseudo trace consists of a set of unrelated pseudo locations.
    \item \textbf{Aggregate Mobility IID~\cite{shokri2011quantifying}:} we synthesize each pseudo location independently and identically from the aggregate mobility profile $\overline{\pi}$. Similarly, the pseudo trace consists of a set of unrelated pseudo locations.
    \item \textbf{Random Walk on Aggregate Mobility~\cite{you2007protecting}:} we synthesize the pseudo location sequence by continuously sampling the next location following the aggregate transition probability distribution $\overline{p}$.
\end{itemize}


The experiment is designed within the context of the Larger scenario, and we deploy the experiments with three different numbers of pseudo trajectories: 1, 5, and 10. Note that the macroscopic model is removed since it will largely surmount the performance decline caused by the privacy mechanisms, as we will discuss in Section ~\ref{par:ablation}.

\begin{figure}[t!]
    \centering
    \subfigure[AUC]
    {\includegraphics[width=.4\textwidth]{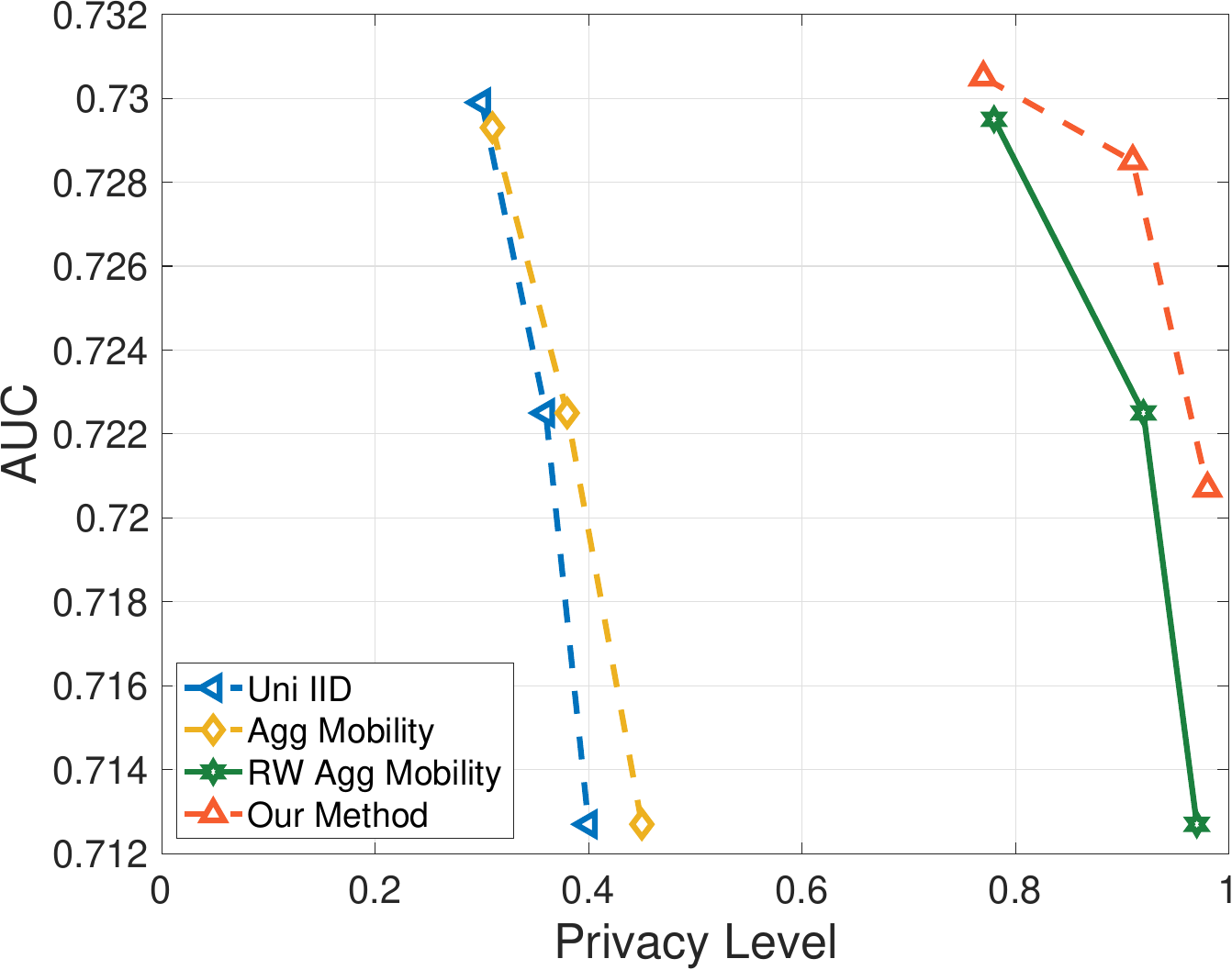}}
    \hspace{1.0cm}
    \subfigure[F1-score]
    {\includegraphics[width=.4\textwidth]{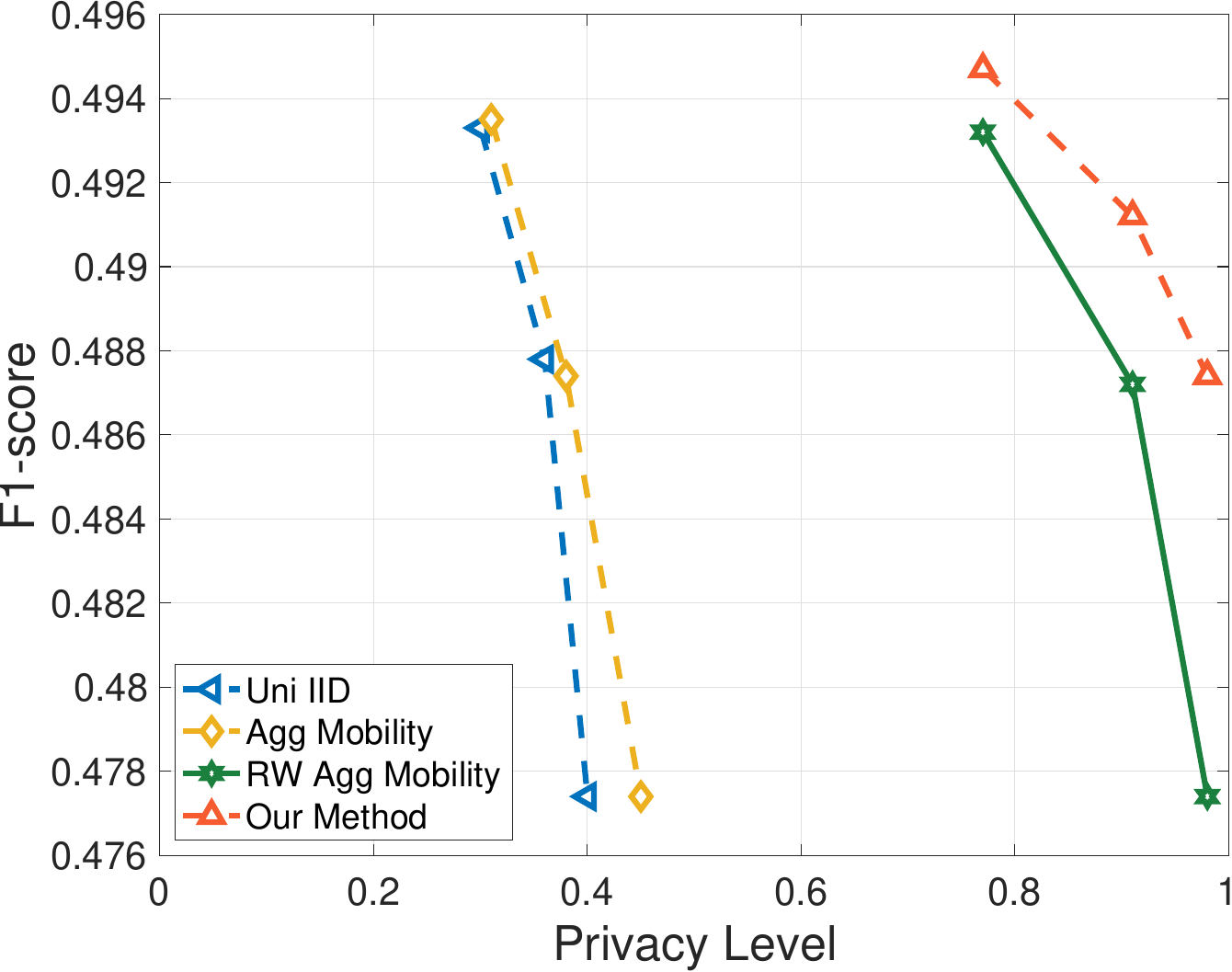}}
    \caption{The trade-off between location privacy and model performance for different synthetic location generation techniques.}\label{fig:privacy-utility}
\end{figure}

In Figure~\ref{fig:privacy-utility}, we illustrate the trade-off between user privacy and model utility for the aforementioned pseudo location generation algorithms. Results show that our method clearly outperforms all the existing techniques, especially the random strategies. For the case of random walk on aggregate mobility, the privacy level against the tracking attack is close to what we achieve, due to the similarity with our method in geographic semantic. But our method achieves less performance decline compared with it by taking random walk on the epidemic domain.

\subsection{Ablation Study}\label{par:ablation}
To evaluate the effectiveness of the macroscopic model, which is proposed to defend against the performance decay produced by privacy-preserving mechanisms, we perform several ablation experiments by blocking specific modules. In particular, we remove the macroscopic model to assess the performance gain brought by it. Then we further remove specific privacy-preserving modules one by one to estimate their impact on model performance and verify the robustness of \emph{Falcon} under the strict privacy budget. Specifically, we remove the perturbation mechanism and the plausible pseudo location generation method respectively, or simultaneously. 
The results are illustrated in Table~\ref{tab:ablation}, where we deploy five variant models of \emph{Falcon}. Particularly, we remove the macroscopic model and all privacy-preserving techniques and align with the HGNN model. With the above results, we can summarize the findings as follows:
\begin{itemize}[leftmargin=*]
    \item \para{With the enhancement of the macroscopic model, \emph{Falcon} can effectively mitigate the performance decay:} \emph{Falcon} not only overcomes the performance decay but also achieves significantly higher prediction accuracy under the various privacy-preserving mechanisms. Besides, with the integration of the macroscopic model, the performance of \emph{Falcon} keeps robust under various privacy strategies.

    \item \para{Even without the macroscopic model, \emph{Falcon} still strikes better prediction performance compared to baseline methods:} 
    We compare the \emph{Falcon} without macroscopic model with all baseline methods that have been mentioned before. Apparently, \emph{Falcon} with strict privacy guarantees always strikes the best overall performance relative to other baselines without any privacy consideration.  
    The phenomenon demonstrates that although the macroscopic model significantly improves prediction performance, even without it the microscopic model remains an innovative and effective approach compared to existing methods.

\end{itemize}

\begin{table}[]
    \centering
    \caption{Results of ablation study.}
    \begin{tabular}{c||ccc}
    \hlineB{3}
    \multicolumn{1}{c||}{ Scenario } & \multicolumn{3}{c}{ Larger } \\
    \multicolumn{1}{c||}{ Method } & DEP & F1 & AUC \\
    \hhline{=::===} 
    HGNN Model & 0.3389 & 0.5016 & 0.7511 \\
    w/o Macro Model & 0.3279 & 0.4947 & 0.7305 \\
    w/o Pseudo Location & 0.4269 & 0.5779 & 0.8279 \\
    w/o Perturbation Mechanism & 0.4237 & 0.5746 & 0.8270 \\
    w/o Privacy Constrict & 0.4281 & 0.5792 & 0.8290 \\
    \hline Falcon & $\mathbf{0.4219}$ & $\mathbf{0.5729}$ & $\mathbf{0.8257}$ \\
    \hlineB{3}
    \end{tabular}
    \label{tab:ablation}
\end{table}

\subsection{Hyperparameters Study}
In the actual scenario, it is impossible for individuals to continuously upload their trajectories and report their infection states. Furthermore, different reporting rates can be attributed to various factors, such as government policies and the initiative of the public. Consequently, we conducted experiments to quantify the impact of two sets of hyperparameters, the report rate of infection states $\lambda$, and the proportion of uploaded trajectory points $\eta$, on the performance of infection prediction models.

\begin{figure}[t!]
    \centering
    \subfigure[Reporting frequency of trajectory.]
    {\includegraphics[width=.48\textwidth]{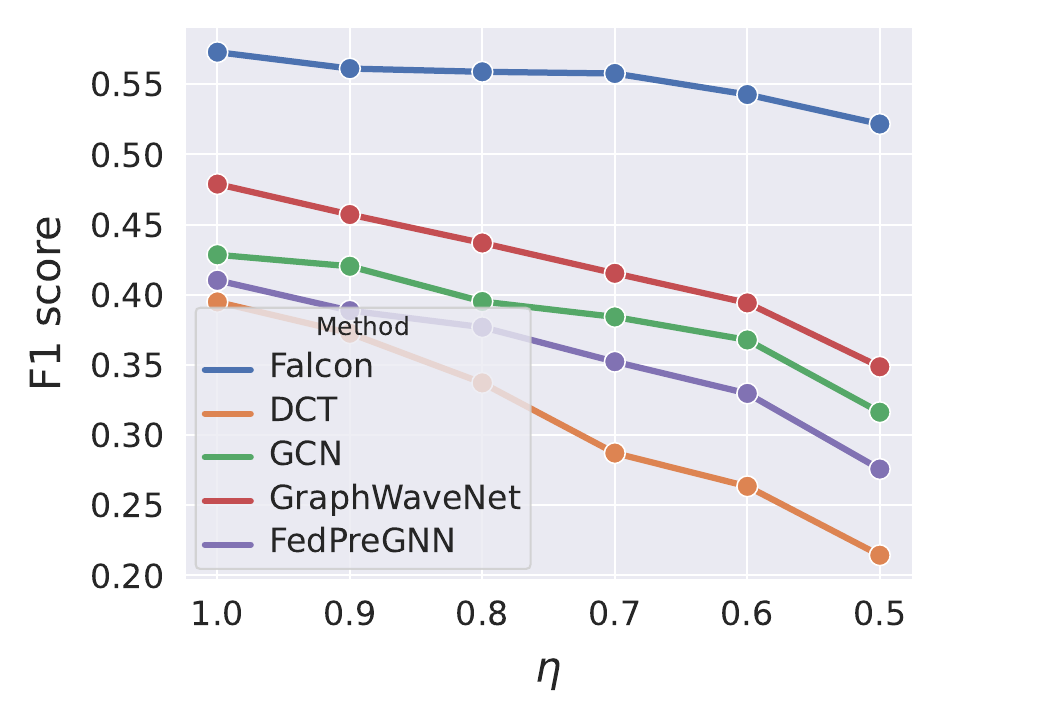}}
    \subfigure[Reporting rate of infection state.]
    {\includegraphics[width=.48\textwidth]{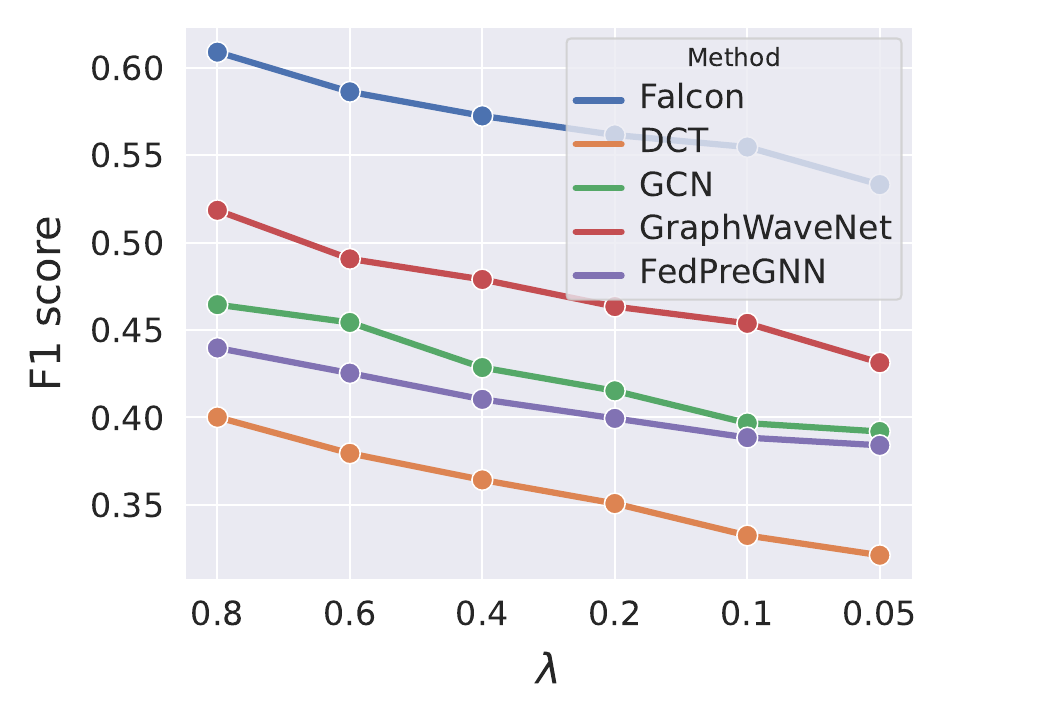}}
    \caption{The infection prediction performance of \textit{Falcon} and representative baselines under different trajectory reporting frequencies and infection state reporting rates.}\label{fig:hyperparameter}
\end{figure}

\para{Trajectory Reporting Frequency:} We conduct the experiment with a set of different reporting frequencies $\eta$ from 1 to 0.5, where 1 represents each individual continuously uploading their complete movement trajectory, while 0.5 represents a probability of 50\% for not uploading the trajectory. The results are summarized in Figure \ref{fig:hyperparameter}(a). We can observe that our proposed \textit{Falcon} consistently achieves the best and most stable prediction performance because it can extract multiscale information of disease transmission through a joint macroscopic model. The performance of baselines shows varying degrees of decline as the reporting frequency decreases, with the DCT method showing a more significant decline. This reveals the high dependence of traditional DCT methods on individual trajectory reporting rates, which aligns with existing research findings \cite{elmokashfi2021nationwide, mancastroppa2021stochastic}. At the same time, this also emphasizes the necessity of combining strong and effective methods such as GNN. 

\para{State Reporting Rate:} Similarly, we carefully selected six different possible reporting rates $\lambda$, ranging from high to low, in order to cover most of the scenarios that may exist in reality. As depicted in Figure \ref{fig:hyperparameter}(b), \textit{Falcon} under the scenarios with lower reporting rates still strikes better prediction performance than baselines with higher reporting rates. Consequently, \textit{Falcon} can maintain relatively good prediction utility, even in cases where individuals are not actively reporting their infection status.

%% file: 2.relatedwork.tex
\section{Related Works}\label{sec::related}

\subsection{Federated Graph Machine Learning (FGML)}
With the rapid development of machine learning and the emerging requirement of privacy protection~\cite{li2020review}, federated learning is proposed to train the global model with massive distributed clients~\cite{konevcny2016federated}, which guarantees user privacy by keeping data in local devices. Further, some researchers introduce differential privacy into federated learning~\cite{abadi2016deep, long2023decentralized} against the inference attack to model gradients~\cite{hitaj2017deep, melis2019exploiting}. In contrast, some works~\cite{wu2022federated, meng2021crossnode, he2022spreadgnn, chen2021fedgraph} consider integrating federated learning with graph neural networks~\cite{kipf2017semi} to develop a federated graph machine learning (FGML) framework. Although the existing works provide a promising paradigm of FGML for multi-scenario applications (e.g., recommendation system~\cite{wu2022federated, liu2022federated}, traffic prediction~\cite{meng2021crossnode} and molecular prediction~\cite{he2022spreadgnn}), there are some critical challenges need to be addressed. Including cross-client missing information~\cite{chen2021fedgraph}, privacy leakage of graph structures~\cite{liu2022federated} and performance decline~\cite{wu2022federated}. To tackle the above challenges, we propose a spatio-temporal hypergraph construction method and separate the propagation procedures of hypergraph into distributed for precise individual-level infection prediction. Besides, a macroscopic model is introduced to address the performance decline.

\subsection{Individual-level Infection Prediction}
Conventional individual-level infection prediction is deployed by digital contact tracing (DCT)~\cite{grekousis2021digital}. However, DCT failed to provide a precise risk of individual being infected in practical~\cite{davis2021contact}, as existing DCT techniques only predict the potential positive cases by deploying cross-check with the trace of confirmed cases~\cite{rambhatla2022toward, da2021react, kato2021pct, peng2021p2b}, the complex interaction among massive individuals are not modeled well~\cite{cebrian2021past}. 

Recently, massive researches investigate the epidemiology domain with the help of GNN, in which some works \cite{gao2021stan,wang2022causalgnn,panagopoulos2021transfer} utilize GNN to embed spatial signals from disease dynamics for achieving more accurate infection prediction in the region level. \citeauthor{gao2021stan} \cite{gao2021stan} introduce a spatio-temporal attention network that fuses recurrent neural networks (RNN) and graph attention networks (GAT) to simultaneously extract geographic trends and temporal patterns during disease transmission. To account for the limited amount of data in some countries,  \citeauthor{panagopoulos2021transfer} \cite{panagopoulos2021transfer} propose a method based on model-agnostic meta-learning to transfer the GNN-based infection prediction model from one country to another where limited data is available. \citeauthor{wang2022causalgnn} \cite{wang2022causalgnn} design a causal module on GNN to further capture the causal temporal signals in the pandemic for better performance.
However, these methods only model interactions at the region level, which fail to capture the intercorrelation among a large number of individuals and accurately assess the infection risk for each person. Thus, other works~\cite{murphy2021deep, tomy2022estimating, feng2022contact} explore employing GNN in individual-level tasks and achieve relatively well performance. ~\citeauthor{murphy2021deep}~\cite{murphy2021deep} utilize a GNN model to learn the contagion dynamics on complex networks and exploit the percolation and phase transitions in the epidemic. 
~\citeauthor{tomy2022estimating}~\cite{tomy2022estimating} propose an individual-level infection prediction model that employs GNN to capture the contact among individuals. ~\citeauthor{feng2022contact}~\cite{feng2022contact} utilize GNN and reinforcement learning to design an individual-level epidemic control agent for searching efficient intervention strategies. Nevertheless, there are some limitations of these works as follows. First, they only model the interactions among individuals, but omit the high-order interactions between individuals and locations. Second, all existing works can not address the privacy concern of using mobility data. In this work, we proposed a novel federated graph learning method for infection prediction, namely \emph{Falcon}, to capture the high-order interactions within a constricted privacy budget.